\documentstyle[epsfig,12pt]{article}
\newcommand{\Z}{{\sf Z \!\!\! Z}}

\newcommand{\up}{\uparrow}
\newcommand{\down}{\downarrow}
\newcommand{\Sign}{\mbox{Sign}}
\makeatletter
\@addtoreset{equation}{section}
\makeatother

\title{Meron-Cluster Simulation of Quantum Spin Ladders in a Magnetic Field
\footnote{This work is supported in part by funds provided by the U.S.
Department of Energy (D.O.E.) under cooperative research agreements
DE-FC02-94ER40818 and DE-FG02-96ER40949.}}
\author{S. Chandrasekharan$^a$, B. Scarlet$^b$ and U.-J. Wiese$^b$
\\ \\
$^a$ Department of Physics, Duke University, \\
Durham, North Carolina 27708 \\
$^b$ Center for Theoretical Physics, Laboratory for Nuclear Science \\
and Department of Physics, Massachusetts Institute of Technology, \\
Cambridge, Massachusetts 02139 \\
\\
DUKE-TH-99-197, MIT-CTP-2904}

\begin{document}
\maketitle
\begin{abstract} \normalsize

Numerical simulations of numerous quantum systems suffer from the notorious 
sign problem. Meron-cluster algorithms lead to an efficient solution of sign 
problems for both fermionic and bosonic models. Here we apply the meron
concept to quantum spin systems in an arbitrary external magnetic field, in  
which case standard cluster algorithms fail. As an example, we simulate 
antiferromagnetic quantum spin ladders in a uniform external magnetic field 
that competes with the spin-spin interaction. The numerical results are in
agreement with analytic predictions for the magnetization as a function of the
external field.

\end{abstract}
 
\maketitle

\newpage

\section{Introduction}

Numerical simulations of quantum systems can suffer from the notorious sign 
problem. Well-known examples of systems where this problem has hindered 
progress are fermions in more than one spatial dimension, quantum 
antiferromagnets on non-bipartite lattices, as well as some quantum spin 
systems in an external magnetic field. For these systems the Boltzmann factor 
of a configuration in the path integral can be negative and hence cannot be 
interpreted as a probability. When the sign of the Boltzmann factor associated
with the configuration is incorporated in measured observables, the 
fluctuations in the sign give rise to dramatic cancellations. In particular, 
for large systems at low temperatures this leads to relative statistical errors
that are exponentially large in both the volume and the inverse temperature. As
a consequence, it is impossible in practice to study such systems with standard
numerical methods. 

Recently, some severe fermion sign problems have been solved with a 
meron-cluster algorithm \cite{Cha99a,Cha99b,Cha99c,Cox99}. Such an algorithm 
was originally developed to solve the sign problem associated with the phase 
$\exp(i\theta Q)$ of the Boltzmann weight in the $(1+1)$-d $O(3)$ symmetric 
quantum field theory at vacuum angle $\theta = \pi$ \cite{Bie95}. Here $Q$ is 
the topological charge that measures the winding number of a configuration and 
that is carried by the classical instanton solutions. The solution of the sign 
problem is based on the fact that the flip of certain clusters changes the 
topological charge by one and hence changes the sign $(-1)^Q$ of the 
configuration. Such clusters can be associated with half a unit of topological 
charge and were referred to as merons \footnote{The term meron is traditionally
used to denote a half-instanton.}. Thus meron clusters are used to identify two
configurations with the same weight but opposite signs. Remarkably, this 
property of a meron is applicable to clusters in a variety of models with sign 
problems. Hence we generally use the term meron to denote clusters whose flip 
changes the sign of a configuration without changing its weight. In this paper 
we extend the meron concept to quantum spin systems in a magnetic field.

As an example of physical interest we consider antiferromagnetic quantum spin 
ladders in an external uniform magnetic field that competes with the spin-spin 
interaction. The competing interactions lead to severe problems in numerical 
simulations. The most efficient algorithms for simulating quantum spin systems 
in the absence of an external magnetic field are cluster algorithms 
\cite{Wie92}, in particular the loop algorithm \cite{Eve93,Wie94,Eve97} which 
can be implemented directly in the Euclidean time continuum \cite{Bea96}. The 
idea behind this algorithm is to decompose a spin configuration into clusters 
which can be flipped independently with probability $1/2$. This procedure leads
to large collective moves through configuration space and eliminates critical 
slowing down. In the presence of a magnetic field pointing along the 
quantization axis of the spins, the $\Z(2)$ symmetry that allows the clusters 
to be flipped with probability $1/2$ is explicitly broken. As a consequence, at
low temperatures or for large values of the magnetic field some clusters can 
only be flipped with very small probability and the algorithm becomes 
inefficient. An interesting alternative to the loop algorithm is the so-called 
worm algorithm \cite{Pro96} which has produced high-precision results for 
quantum spin chains in a magnetic field \cite{Kas99}. The worm algorithm 
explores an enlarged configuration space and, unlike the loop algorithm, seems 
not to suffer from exponential slowing down at least for 1-d spin chains in a 
magnetic field. We are not aware of any applications of the worm algorithm to 
higher-dimensional spin systems, and it would be interesting to see if this 
algorithm then still works efficiently. Another method based on stochastic 
series expansion techniques has been applied successfully to 2-d quantum spin 
systems in a magnetic field \cite{San99}. An alternative strategy using cluster
algorithms is to choose the spin quantization axis perpendicular to the 
direction of the magnetic field. Then all clusters can still be flipped with 
probability $1/2$. This procedure indeed leads to an efficient algorithm for 
ferromagnets in an external uniform magnetic field \footnote{This was pointed 
out to one of the authors by H. G. Evertz and M. Troyer some time ago.}. 
Unfortunately, for antiferromagnets, i.e. when the magnetic field competes with
the spin-spin interaction, this formulation of the problem leads to a very 
severe sign problem. In this paper we show how this sign problem can be solved 
completely using a meron-cluster algorithm. This method allows us to simulate 
quantum spin systems in an arbitrary magnetic field, in which case the standard
loop algorithm fails. Indeed, we show in this paper that, at low temperatures 
or for large values of the magnetic field, the meron-cluster algorithm is far 
more efficient than the loop algorithm. Since the meron-cluster algorithm has 
all the advantages of a cluster algorithm, including improved estimators for 
various physical observables, we expect it to be more efficient than the worm 
algorithm. A detailed comparison of the worm algorithm, the stochastic series 
expansion technique and the meron-cluster algorithm would be interesting for 
future studies.

Antiferromagnetic quantum spin ladders are interesting systems that interpolate
between 1-dimensional spin chains and 2-dimensional spin systems. The 2-d 
systems may become high-temperature superconductors after doping. As reviewed 
in \cite{Dag96}, Haldane's conjecture \cite{Hal83} has been generalized to 
quantum spin ladders: ladders consisting of an odd number of transversely 
coupled spin $1/2$ chains are gapless, while ladders consisting of an even 
number of chains have a gap \cite{Khv94,Whi94,Sie96}. Correlations in quantum 
spin ladders were studied numerically in \cite{Gre96,Syl97}. When the even 
number of coupled chains is increased, the gap decreases exponentially. In the 
limit of a large even number of transversely coupled chains one approaches the 
continuum limit of the 2-d classical $O(3)$ model \cite{Cha96}, which can be 
viewed as a $(1+1)$-d relativistic quantum field theory. When the quantum spin 
ladder is placed in an external uniform magnetic field $B$, the corresponding 
quantum field theory is the $(1+1)$-d $O(3)$ model with chemical potential 
$\mu = B/c$ (where $c$ is the spin-wave velocity). Using Bethe ansatz 
techniques, this theory has been solved exactly. Translating the field 
theoretic results back into the language of condensed matter physics, we obtain
an analytic expression for the magnetization of a quantum spin ladder 
consisting of a large even number of transversely coupled spin $1/2$ chains. 
The numerical results obtained with the meron-cluster algorithm are in 
agreement with the analytic predictions.

The rest of this paper is organized as follows. Section 2 contains a general
discussion of the nature of the sign problem and illustrates the basic ideas
behind the meron-cluster algorithm. In section 3 the path integral 
representation for quantum magnets is derived. Section 4 contains a description
of the cluster algorithm for quantum spin systems and section 5 introduces the
meron concept and discusses the solution of the sign problem in
detail. In section 6 the physics of antiferromagnetic quantum spin
ladders is discussed theoretically and the results of numerical
simulations are presented in section 7. Finally, section 8 contains
our conclusions. 

\section{Nature of the Sign Problem and the Meron Concept}

When antiferromagnetic quantum spin systems in an external magnetic field are 
simulated numerically, a severe sign problem arises. The general nature of this
problem was discussed in \cite{Cha99a}. For the convenience of the reader and 
to make this paper self-contained we reproduce this discussion here. We 
consider the partition function of some quantum system written as a path 
integral
\begin{equation}
Z = \sum_s \Sign[s] \exp(-S[s]),
\end{equation}
over configurations $s$ with a Boltzmann weight of magnitude
$\exp(-S[s])$ and $\Sign[s] = \pm 1$. Here $S[s]$ is the action of a
system with a positive Boltzmann weight and a modified partition
function 
\begin{equation}
Z' = \sum_s \exp(-S[s]),
\end{equation}
which does not suffer from the sign problem. An observable $O[s]$ of the
original system is obtained in a simulation of the modified model as
\begin{equation}
\langle O \rangle = \frac{1}{Z} \sum_s O[s] \Sign[s] \exp(-S[s]) =
\frac{\langle O \Sign \rangle}{\langle \Sign \rangle}.
\end{equation}
The average sign in the modified ensemble is given by
\begin{equation}
\langle \Sign \rangle = \frac{1}{Z'}{\sum_s \Sign[s] \exp(-S[s])} = 
\frac{Z}{Z'} = \exp(- \beta V \Delta f).
\end{equation}
The expectation value of the sign is exponentially small in both the volume $V$
and the inverse temperature $\beta$ because the difference $\Delta f$ between 
the free energy densities of the original and the modified systems is always 
positive. Hence the original expectation value $\langle O \rangle$ --- 
although of order one --- is calculated as the ratio of two exponentially small
modified expectation values $\langle O \Sign \rangle$ and 
$\langle \Sign \rangle$ which are very small compared to the statistical error 
and therefore impossible to measure in practice. This difficulty is the origin 
of the sign problem. We can estimate the statistical error of the average sign 
in an ideal simulation of the modified ensemble which generates $N$ completely 
uncorrelated configurations as
\begin{equation}
\frac{\Delta \Sign}{\langle \Sign \rangle} = 
\frac{\sqrt{\langle \Sign^2 \rangle - \langle \Sign \rangle^2}}
{\sqrt{N} {\langle \Sign \rangle}} = \frac{\exp(\beta V \Delta f)}{\sqrt{N}}.
\end{equation}
The last equality results from taking the large $\beta V$ limit and by using
$\Sign^2 = 1$. To determine the average sign with sufficient accuracy one needs
on the order of $N = \exp(2 \beta V \Delta f)$ measurements. For large volumes 
and small temperatures this is infeasible. 

A naive meron-cluster algorithm simply matches any contribution of $-1$ with 
another contribution of $1$ to give $0$, such that only a few unmatched 
contributions of $1$ remain. Then effectively $\Sign = 0,1$ and hence 
$\Sign^2 = \Sign$. This step reduces the relative statistical error to
\begin{equation}
\frac{\Delta \Sign}{\langle \Sign \rangle} =
\frac{\sqrt{\langle \Sign \rangle - \langle \Sign \rangle^2}}
{\sqrt{N'} {\langle \Sign \rangle}} = 
\frac{\exp(\beta V \Delta f/2)}{\sqrt{N'}}.
\end{equation}
One gains an exponential factor in statistics, but one still needs to generate 
$N' = \sqrt{N} = \exp(\beta V \Delta f)$ independent configurations in order to
accurately determine the average sign. Since the probability to measure a 
contribution of $1$ is $\langle \Sign \rangle$, one generates exponentially 
many contributions of $0$ before one encounters a contribution of $1$. Thus 
this step alone solves only one half of the sign problem and would at most 
allow us to double the space-time volume $\beta V$.\footnote{The fact that an 
improved estimator alone cannot solve the sign problem was pointed out to one 
of the authors by H. G. Evertz a long time ago.} In a second step involving a 
Metropolis decision, the full meron-cluster algorithm ensures that 
contributions of $0$ are suppressed. This step saves another exponential factor
in statistics and solves the other half of the sign problem.

A central idea behind our algorithm is to decompose a configuration of quantum 
spins into clusters which can be flipped independently. A cluster flip changes 
spin up to spin down and vice versa. The clusters can be characterized by their
effect on the sign. Clusters whose flip changes the sign are referred to as 
merons. Using an improved estimator, any configuration that contains merons 
contributes $0$ to the average sign, since the flip of a meron-cluster leads to
a sign change and hence to an explicit cancellation of the contributions of two
configurations with the same weight but opposite $\Sign[s] = \pm 1$. Thus only 
the configurations without merons make non-vanishing contributions. If the 
clusters can be flipped such that one reaches a reference configuration with a 
positive sign, configurations without merons always have $\Sign[s] = 1$. This 
solves one half of the sign problem. 

The other half of the problem is solved using a reweighting method that 
suppresses multi-meron configurations. In order to perform this essential step,
all observables must be measured using improved estimators. Fortunately, this 
is possible for a variety of physical quantities. For example, the uniform and 
staggered magnetization get non-zero contributions only from the zero-meron
sector, while the associated susceptibilities receive non-zero contributions 
from the two-meron sector as well. Since in this article we will be interested 
in the uniform magnetization only, we can completely eliminate all 
configurations containing meron-clusters. In principle, this restriction can be
achieved with a simple reject step. In this way an exponentially large portion 
of the configuration space of the modified model is eliminated. Configurations 
with merons would contribute in the modified model that does not include the 
sign, but they do not contribute to the observables of the original model with 
the sign. Eliminating the exponentially large multi-meron sector of the 
configuration space saves an exponential factor in statistics and solves the 
other half of the sign problem. In order to improve the autocorrelations of the
algorithm it is advantageous not to eliminate multi-meron configurations 
completely. In practice, a Metropolis step is used instead to suppress strongly
the incidence of multi-meron configurations while still allowing occasional 
visits of sectors with a small number of merons.

\section{Path Integral for Quantum Magnets}

We consider a system of quantum spins $1/2$ on a $d$-dimensional cubic lattice 
with site label $x$ and with periodic spatial boundary conditions. In 
particular, we will be interested in ladder systems on a 2-d rectangular 
lattice of size $L \times L'$ with $L \gg L'$. The spins located at the sites 
$x$ are described by operators $S^i_x$ 
with the usual commutation relations
\begin{equation}
[S_x^i,S_y^j] = i \delta_{xy} \epsilon_{ijk} S_x^k.
\end{equation}
The Hamilton operator 
\begin{equation}
H = \sum_{x,i} [J (S_x^1 S_{x+\hat i}^1 + S_x^2 S_{x+\hat i}^2) +
J' S_x^3 S_{x+\hat i}^3] - \sum_x B_x S_x^1
\end{equation}
couples the spins at the lattice sites $x$ and $x+\hat i$, where $\hat i$ is a 
unit-vector in the $i$-direction. The cases $J' = J < 0$ and $J' = J > 0$ 
correspond to the ferro- and antiferromagnetic quantum Heisenberg model 
respectively, while $J' = 0$ corresponds to the quantum XY-model. We consider a
magnetic field $B_x$ that points in the $1$-direction and has an arbitrary 
$x$-dependence. Later it will be important that the magnetic field is 
perpendicular to the spin quantization axis, which we will choose in the 
$3$-direction. As a specific example we will consider an antiferromagnetic 
isotropic quantum spin ladder (i.e. $J' = J > 0$) in a uniform magnetic field 
$B_x = B$.

To derive a path integral representation of the partition function we decompose
the Hamilton operator into $2d+1$ terms
\begin{equation}
H = H_1 + H_2 + ... + H_{2d+1}.
\end{equation}
The various terms take the forms
\begin{equation}
H_i = \!\!\! \sum_{\stackrel{x = (x_1,x_2,...,x_d)}{x_i even}} \!\!\! h_{x,i},
\, 
H_{i+d} = \!\!\! \sum_{\stackrel{x = (x_1,x_2,...,x_d)}{x_i odd}} \!\!\! 
h_{x,i}, \,
H_{2d+1} = \!\!\! \sum_{\stackrel{x = (x_1,x_2,...,x_d)}{}} \!\!\! h_x,
\end{equation}
with
\begin{equation}
h_{x,i} = J (S_x^1 S_{x+\hat i}^1 + S_x^2 S_{x+\hat i}^2) +
J' S_x^3 S_{x+\hat i}^3, \, h_x = - B_x S_x^1.
\end{equation}
Note that the individual contributions to a given $H_i$ commute with each 
other, but two different $H_i$ do not commute. Using the Trotter-Suzuki formula
we express the partition function as
\begin{equation}
Z = \mbox{Tr} [\exp(- \beta H)] = 
\lim_{M \rightarrow \infty} \!\!\! \mbox{Tr} 
[\exp(- \beta H_1) \exp(- \beta H_2) ... \exp(- \beta H_{2d+1})]^M.
\end{equation}
We have introduced $(2d+1)M$ Euclidean time slices with $\epsilon = \beta/M$ 
as the lattice spacing in the Euclidean time direction. We insert complete 
sets of eigenstates $|\up\rangle$ and $|\down\rangle$ with eigenvalues 
$S_x^3 = \pm 1/2$ between the factors $\exp(- \beta H_i)$. The transfer matrix 
is a product of factors
\begin{equation}
\label{transfer1}
\exp(- \epsilon h_{x,i}) = \left(\begin{array}{cccc}
\exp(- \epsilon J'/2) & 0 & 0 & 0 \\ 
0 & \cosh(\epsilon J/2) & - \sinh(\epsilon J/2) & 0 \\ 
0 & - \sinh(\epsilon J/2) & \cosh(\epsilon J/2) & 0 \\ 
0 & 0 & 0 & \exp(- \epsilon J'/2) \end{array} \right)
\end{equation}
(here we have dropped an irrelevant overall prefactor $\exp(\epsilon J'/4)$) 
and
\begin{equation}
\label{transfer2}
\exp(- \epsilon h_x) = \left(\begin{array}{cc}
\cosh(\epsilon B_x/2) & \sinh(\epsilon B_x/2) \\ 
\sinh(\epsilon B_x/2) & \cosh(\epsilon B_x/2) \end{array} \right).
\end{equation}
The $4 \times 4$ matrix of eq.(\ref{transfer1}) is represented in the basis 
$|\up\up\rangle$, $|\up\down\rangle$, $|\down\up\rangle$ and 
$|\down\down\rangle$ of two sites $x$ and $x+\hat i$, and the $2 \times 2$ 
matrix of eq.(\ref{transfer2}) is in the basis $|\up\rangle$ and
$|\down\rangle$ of the site $x$.

The partition function is now expressed as a path integral
\begin{equation}
Z = \sum_s \Sign[s] \exp(- S[s]),
\end{equation}
over configurations of spins $s(x,t) = \up, \down$ on a $(d+1)$-dimensional 
space-time lattice of points $(x,t)$. The Boltzmann factor
\begin{eqnarray}
\exp(- S[s])&=&\!\!\!\!\! \prod_{\stackrel{x = (x_1,x_2,...,x_d)}
{x_1 even, t = 2dp}} \!\!\!\!\!\!\!\!
\exp\{- S[s(x,t),s(x+\hat 1,t),s(x,t+1),s(x+\hat 1,t+1)]\} \nonumber \\
&\times&\!\!\!\!\! \prod_{\stackrel{x = (x_1,x_2,...,x_d)}
{x_2 even, t = 2dp+1}} \!\!\!\!\!\!\!\!
\exp\{- S[s(x,t),s(x+\hat 2,t),s(x,t+1),s(x+\hat 2,t+1)]\} ... \nonumber \\
&\times&\!\!\!\!\! \prod_{\stackrel{x = (x_1,x_2,...,x_d)}
{x_d even, t = 2dp+d-1}} \!\!\!\!\!\!\!\!
\exp\{- S[s(x,t),s(x+\hat d,t),s(x,t+1),s(x+\hat d,t+1)]\} \nonumber \\
&\times&\!\!\!\!\! \prod_{\stackrel{x = (x_1,x_2,...,x_d)}
{x_1 odd, t = 2dp+d)}} \!\!\!\!\!\!\!\!
\exp\{- S[s(x,t),s(x+\hat 1,t),s(x,t+1),s(x+\hat 1,t+1)]\} \nonumber \\
&\times&\!\!\!\!\! \prod_{\stackrel{x = (x_1,x_2,...,x_d)}
{x_2 odd, t = 2dp+d+1}} \!\!\!\!\!\!\!\!
\exp\{- S[s(x,t),s(x+\hat 2,t),s(x,t+1),s(x+\hat 2,t+1)]\} ... \nonumber \\
&\times&\!\!\!\!\! \prod_{\stackrel{x = (x_1,x_2,...,x_d)}
{x_d odd, t = 2d(p+1)-1}} \!\!\!\!\!\!\!\!
\exp\{- S[s(x,t),s(x+\hat d,t),s(x,t+1),s(x+\hat d,t+1)]\} \nonumber \\ \
&\times&\!\!\!\!\! \prod_{\stackrel{x = (x_1,x_2,...,x_d)}
{t = 2d(p+1)}} \!\!\!\!\!\!\!\!
\exp\{- S[s(x,t),s(x,t+1)]\} \nonumber \\ \
\end{eqnarray}
(with $p = 0,1,...,M-1$) is a product of space-time plaquette contributions 
with
\begin{eqnarray}
\label{Boltzmann1}
&&\exp(- S[\up,\up,\up,\up]) = \exp(- S[\down,\down,\down,\down]) = 
\exp(- \epsilon J'/2), 
\nonumber \\
&&\exp(- S[\up,\down,\up,\down]) = \exp(- S[\down,\up,\down,\up]) = 
\cosh(\epsilon J/2),
\nonumber \\
&&\exp(- S[\up,\down,\down,\up]) = \exp(- S[\down,\up,\up,\down]) = 
\sinh(\epsilon |J|/2),
\end{eqnarray}
and time-like bond contributions
\begin{eqnarray}
\label{Boltzmann2}
&&\exp(- S[\up,\up]) = \exp(- S[\down,\down]) = \cosh(\epsilon B_x/2),
\nonumber \\
&&\exp(- S[\up,\down]) = \exp(- S[\down,\up]) = \sinh(\epsilon |B_x|/2).
\end{eqnarray}

The sign of a configuration, $\Sign[s]$, also is a product of space-time 
plaquette contributions
$\mbox{Sign}[s(x,t),s(x+\hat i,t),s(x,t+1),s(x+\hat i,t+1)]$ with
\begin{eqnarray}
&&\mbox{Sign}[\up,\up,\up,\up]) = \mbox{Sign}[\up,\down,\up,\down]) = 
\mbox{Sign}[\down,\up,\down,\up]) = \mbox{Sign}[\down,\down,\down,\down]) = 1 
\nonumber \\
&&\mbox{Sign}[\up,\down,\down,\up]) = \mbox{Sign}[\down,\up,\up,\down]) = 
- \mbox{sign}(J),
\end{eqnarray}
and time-like bond contributions
\begin{eqnarray}
&&\mbox{Sign}[\up,\up]) = \mbox{Sign}[\down,\down]) = 1 \nonumber \\
&&\mbox{Sign}[\up,\down]) = \mbox{Sign}[\down,\up]) = \mbox{sign}(B_x).
\end{eqnarray}
Figure 1 shows two spin configurations in $(1+1)$ dimensions. The first
configuration is completely antiferromagnetically ordered and has 
$\Sign[s] = 1$. The second configuration contains one interaction plaquette
with configuration $[\down,\up,\up,\down]$ which contributes 
$\Sign[\down,\up,\up,\down] = - 1$ for $J > 0$. In addition, there are two 
time-like interaction bonds with configurations $[\down,\up]$ and 
$[\up,\down]$. When $B_x > 0$ these contribute $\Sign[\down,\up] = 
\Sign[\up,\down] = 1$, such that the whole configuration has $\Sign[s] = - 1$.
\begin{figure}[htb]
\hbox{
\hspace{2.7cm}
${\rm Sign[s]}=1$
\hspace{5.0cm}
${\rm Sign[s]}=-1$
}  
\begin{center}
\hbox{
\epsfig{file=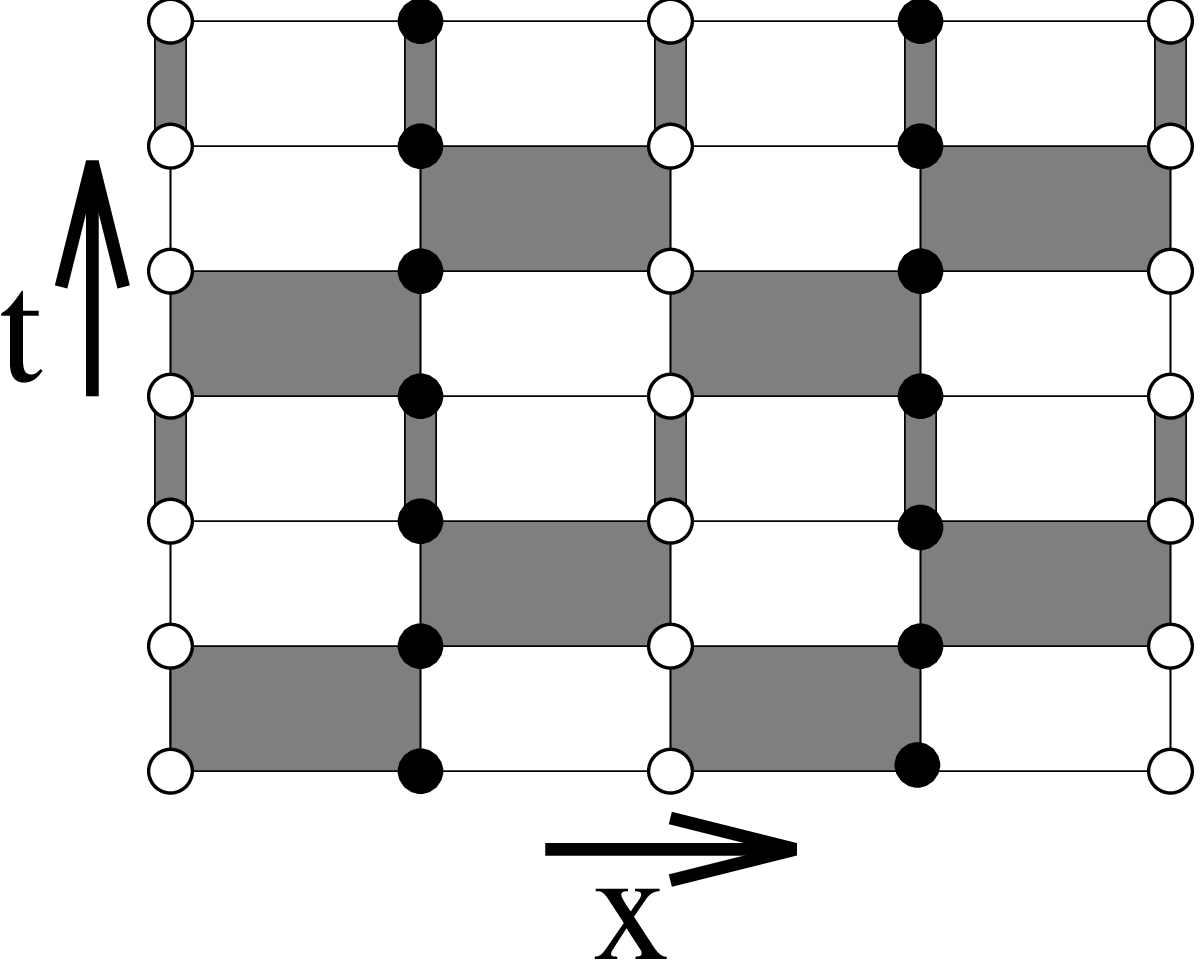,
width=4.5cm,angle=270,
bbllx=0,bblly=0,bburx=225,bbury=337}
\hspace{1.0cm}
\epsfig{file=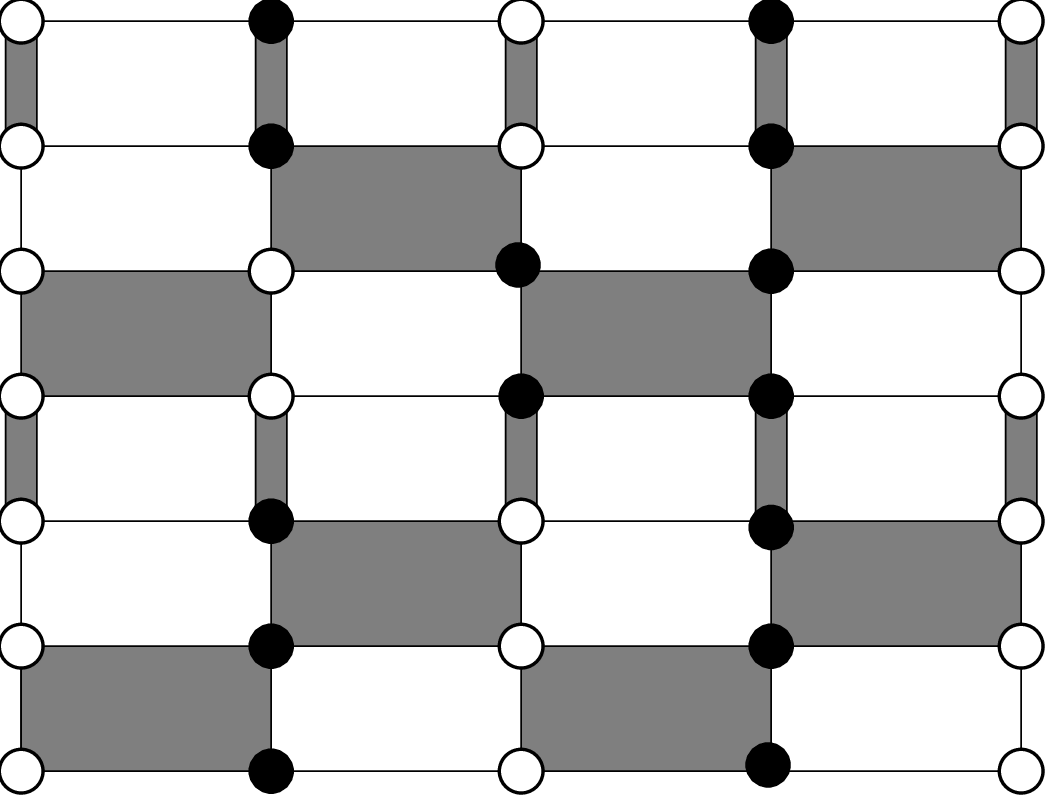,
width=4.5cm,angle=270,
bbllx=0,bblly=0,bburx=225,bbury=337}
}  
\end{center}
\caption{\it Two spin configurations in $(1+1)$ dimensions. The shaded
plaquettes and time-like bonds carry the interaction. Filled dots represent 
spin up and open circles represent spin down. For an antiferromagnetic spin 
chain in a positive magnetic field $B_x > 0$, the second configuration has 
$\Sign[s] = - 1$.}
\end{figure}

The central observable of our study is the uniform magnetization
\begin{equation}
\label{magnetization}
M^i = \sum_x S_x^i.
\end{equation}
The expectation value of the magnetization $\langle M^1 \rangle$ in the 
direction of the magnetic field is non-zero, while $\langle M^2 \rangle = 
\langle M^3 \rangle = 0$.

It should be noted that the path integral can be simplified for isotropic 
ferromagnets ($J' = J < 0$) or antiferromagnets ($J' = J > 0$) in a uniform 
magnetic field ($B_x = B$). In that case, the magnetic field interaction 
$B \sum_x S_x^1$ commutes with the rest of the Hamiltonian and can hence be 
concentrated in a single Euclidean time-step. Then one can work with only 
$2dM+1$ time-slices instead of $(2d+1)M$. Instead of $M$ magnetic field 
interaction time-steps with Boltzmann weights $\cosh(\epsilon B/2)$ and 
$\sinh(\epsilon B/2)$ one then has only one such step with Boltzmann weights 
$\cosh(\beta B/2)$ and $\sinh(\beta B/2)$. In the simulations performed in 
this paper we use the method with $(2d+1)M$ time-slices.

\section{Cluster Algorithm for the Modified Model without the Sign Factor}

The meron-cluster algorithm is based on a cluster algorithm for the modified 
model without the sign factor. Quantum spin systems without a sign problem can 
be simulated very efficiently with the loop-cluster algorithm 
\cite{Eve93,Wie94,Eve97}. This algorithm can be implemented directly in the 
Euclidean time continuum \cite{Bea96}, i.e. the Suzuki-Trotter discretization 
is not even necessary. The same is true for the meron-cluster algorithm. Here 
we discuss the algorithm for discrete time. 

The idea behind the algorithm is to decompose a configuration into clusters
which can be flipped independently. Each lattice site belongs to exactly one
cluster. When the cluster is flipped, the spins at all the sites on the cluster
are changed from up to down and vice versa. The decomposition of the lattice
into clusters results from connecting neighboring sites on each space-time 
interaction plaquette or time-like interaction bond according to probabilistic 
cluster rules. A set of connected sites defines a cluster. In this case the 
clusters are open or closed strings. The cluster rules are constructed so as to
obey detailed balance. To show this property we first write the space-time 
plaquette 
Boltzmann factors as
\begin{eqnarray}
\label{cluster1}
&&\!\!\!\exp(- S[s(x,t),s(x+\hat i,t),s(x,t+1),s(x+\hat i,t+1)]) = \nonumber \\
&&\!\!\!A \delta_{s(x,t),s(x,t+1)} \delta_{s(x+\hat i,t),s(x+\hat i,t+1)} +
B \delta_{s(x,t),-s(x+\hat i,t)} \delta_{s(x,t+1),-s(x+\hat i,t+1)} +
\nonumber \\
&&\!\!\!C \delta_{s(x,t),s(x,t+1)} \delta_{s(x+\hat i,t),s(x+\hat i,t+1)}
\delta_{s(x,t),-s(x+\hat i,t)} + 
D \delta_{s(x,t),s(x+\hat i,t+1)} \delta_{s(x+\hat i,t),s(x,t+1)} +
\nonumber \\
&&\!\!\!E \delta_{s(x,t),s(x,t+1)} \delta_{s(x+\hat i,t),s(x+\hat i,t+1)}
\delta_{s(x,t),s(x+\hat i,t)}.
\end{eqnarray}
The various $\delta$-functions specify which sites are connected and thus
belong to the same cluster. It is possible to construct a cluster algorithm
satisfying detailed balance if the coefficients $A,B,...,E$ are all 
non-negative. In that case these coefficients determine the relative 
probabilities for different cluster break-ups of an interaction plaquette. 
\footnote{We thank R. Brower for introducing us to the $\delta$-function 
method.} For example, $A$ determines the probability with which sites are 
connected with their time-like neighbors, while $B$ and $D$ determine the 
probabilities for connections with space-like and diagonal neighbors. The 
quantities $C$ and $E$ determine the probabilities to put all four spins of a 
plaquette into the same cluster. This is possible for plaquette configurations 
$[\up,\down,\up,\down]$ or $[\down,\up,\down,\up]$ with a probability 
proportional to $C$ and for configurations $[\up,\up,\up,\up]$ or
$[\down,\down,\down,\down]$ with a probability proportional to $E$. Inserting 
the expressions from eq.(\ref{Boltzmann1}) one finds
\begin{eqnarray}
\label{balance1}
&&\exp(- S[\up,\up,\up,\up]) = \exp(- S[\down,\down,\down,\down]) = 
\exp(- \epsilon J'/2) = A + D + E, \nonumber \\
&&\exp(- S[\up,\down,\up,\down]) = \exp(- S[\down,\up,\down,\up]) = 
\cosh(\epsilon J/2) = A + B + C, \nonumber \\
&&\exp(- S[\up,\down,\down,\up]) = \exp(- S[\down,\up,\up,\down]) = 
\sinh(\epsilon |J|/2) = B + D.
\end{eqnarray}
For example, the probability to connect the sites with their time-like
neighbors on a plaquette with configuration $[\up,\up,\up,\up]$ or 
$[\down,\down,\down,\down]$ is $A/(A+D+E)$, while the probability to connect
them with their diagonal neighbor is $D/(A+D+E)$. All sites on such a
plaquette are connected together and are hence put into the same cluster with 
probability $E/(A+D+E)$. Similarly, the probability for connecting space-like 
neighbors on a plaquette with configuration $[\up,\down,\down,\up]$ or 
$[\down,\up,\up,\down]$ is $B/(B+D)$ and the probability for connecting
diagonal neighbors is $D/(B+D)$. Similarly, the time-like bond Boltzmann 
factors are expressed as
\begin{equation}
\label{cluster2}
\exp(- S[s(x,t),s(x,t+1)]) = F \delta_{s(x,t),s(x,t+1)} + G.
\end{equation}
The probability to connect spins with their time-like neighbors is $F/(F+G)$.
The spins remain disconnected with probability $G/(F+G)$. Inserting the 
expressions from eq.(\ref{Boltzmann2}) one obtains
\begin{eqnarray}
\label{balance2}
&&\exp(- S[\up,\up]) = \exp(- S[\down,\down]) = \cosh(\epsilon B_x/2) = F + G, 
\nonumber \\
&&\exp(- S[\up,\down]) = \exp(- S[\down,\up]) = \sinh(\epsilon |B_x|/2) = F.
\end{eqnarray}
The cluster rules are illustrated in table 1.
\begin{table}[htb]
% space before first and after last column: 1.5pc
% space between columns: 3.0pc (twice the above)
%\setlength{\tabcolsep}{1.5pc}
% -----------------------------------------------------
% adapted from TeX book, p. 241
%\newlength{\digitwidth} \settowidth{\digitwidth}{\rm 0}
\catcode`?=\active \def?{\kern\digitwidth}
% -----------------------------------------------------
  \begin{center}
    \leavevmode
%    \tiny
    \begin{tabular}{|c|c|c|}
	\hline
	weight&configuration&break-ups\\
	\hline\hline
	$\exp(- \epsilon J'/2)$
	&\hspace{0.4cm}\begin{minipage}[c]{2.5cm}
		\epsfig{file=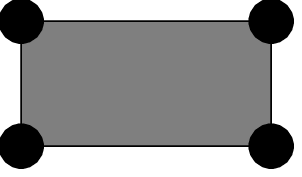,width=1.5cm,angle=270,
		bbllx=-5,bblly=0,bburx=55,bbury=85}
	\end{minipage}
	&\hspace{0.4cm}\begin{minipage}[c]{2.5cm}
		\vbox{\begin{center}
		\epsfig{file=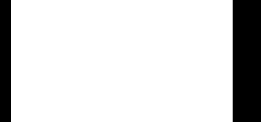,width=1.5cm,angle=270,
		bbllx=-20,bblly=0,bburx=55,bbury=85} 

		\hspace{-0.2cm}A
		\end{center}}
	\end{minipage}
	\hspace{0.4cm}\begin{minipage}[c]{2.5cm}
		\vbox{\begin{center}
		\epsfig{file=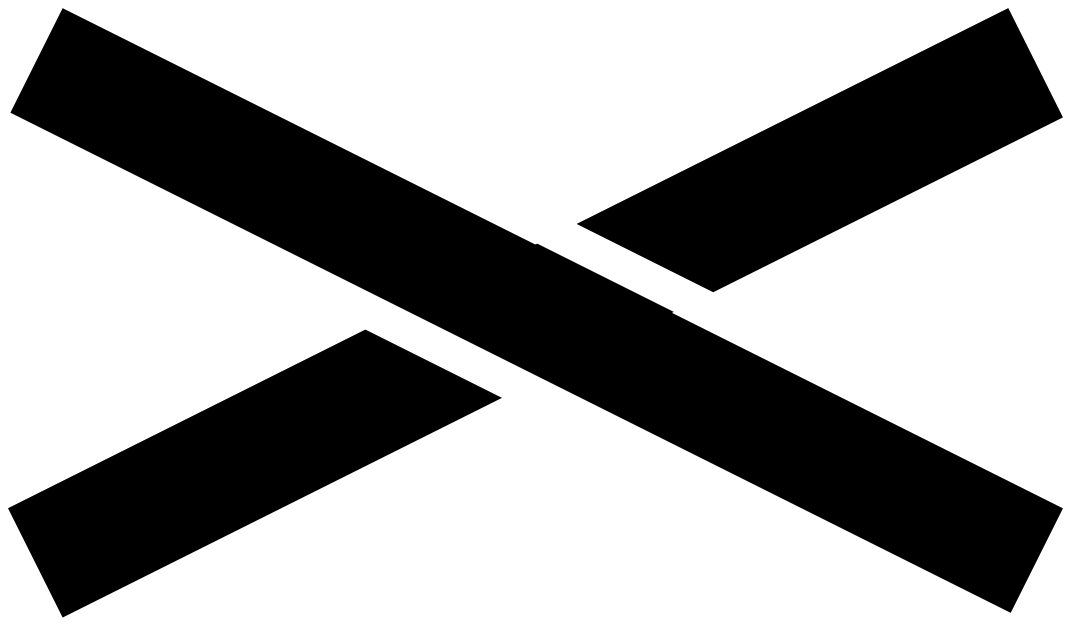,width=1.5cm,angle=270,
		bbllx=-50,bblly=50,bburx=250,bbury=300} 

		\hspace{-0.2cm}D
		\end{center}}
	\end{minipage}
        \hspace{0.4cm}\begin{minipage}[c]{2.5cm}
		\vbox{\begin{center}
		\epsfig{file=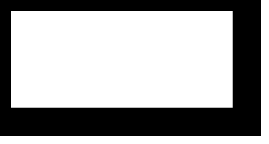,width=1.5cm,angle=270,
		bbllx=-20,bblly=0,bburx=55,bbury=85} 

		\hspace{-0.2cm}E
		\end{center}}
	\end{minipage}
	\\
	\hline
	$\cosh(\epsilon J/2)$
	&\hspace{0.4cm}\begin{minipage}[c]{2.5cm}
		\epsfig{file=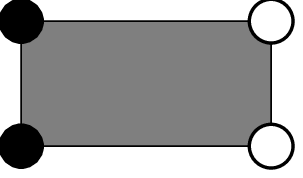,width=1.5cm,angle=270,
		bbllx=-5,bblly=0,bburx=55,bbury=85}
	\end{minipage}
	&\hspace{0.4cm}\begin{minipage}[c]{2.5cm}
		\vbox{\begin{center}
		\epsfig{file=table_1_2.eps,width=1.5cm,angle=270,
		bbllx=-20,bblly=0,bburx=55,bbury=85} 

		\hspace{-0.2cm}A
		\end{center}}
	\end{minipage}
	\hspace{0.4cm}\begin{minipage}[c]{2.5cm}
		\vbox{\begin{center}
		\epsfig{file=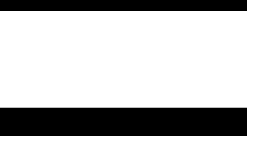,width=1.5cm,angle=270,
		bbllx=-20,bblly=0,bburx=55,bbury=85} 

		\hspace{-0.2cm}B
		\end{center}}
	\end{minipage}
	\hspace{0.4cm}\begin{minipage}[c]{2.5cm}
		\vbox{\begin{center}
		\epsfig{file=table_1_4.eps,width=1.5cm,angle=270,
		bbllx=-20,bblly=0,bburx=55,bbury=85} 

		\hspace{-0.2cm}C
		\end{center}}
	\end{minipage}
	\\
	\hline
	$\sinh(\epsilon |J|/2)$
	&\hspace{0.4cm}\begin{minipage}[c]{2.5cm}
		\epsfig{file=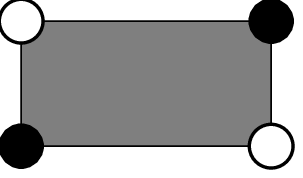,width=1.5cm,angle=270,
		bbllx=-5,bblly=0,bburx=55,bbury=85}
	\end{minipage}
	&\hspace{0.4cm}\begin{minipage}[c]{2.5cm}
		\vbox{\begin{center}
		\epsfig{file=table_3_2.eps,width=1.5cm,angle=270,
		bbllx=-20,bblly=0,bburx=55,bbury=85} 

		\hspace{-0.2cm}B
		\end{center}}
	\end{minipage}
	\hspace{0.4cm}\begin{minipage}[c]{2.5cm}
		\vbox{\begin{center}
		\epsfig{file=table_1_3.eps,width=1.5cm,angle=270,
		bbllx=-50,bblly=50,bburx=250,bbury=300} 

		\hspace{-0.2cm}D
		\end{center}}
	\end{minipage}
	\\
	\hline   
	$\cosh(\epsilon B_x/2)$
	&\hspace{1.4cm}\begin{minipage}[c]{1.5cm}
		\epsfig{file=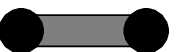,width=1.5cm,angle=270,
		bbllx=-5,bblly=0,bburx=55,bbury=85}
	\end{minipage}
	&\hspace{0.4cm}\begin{minipage}[c]{0.5cm}
		\vbox{\begin{center}
		\epsfig{file=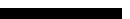,width=1.5cm,angle=270,
		bbllx=-10,bblly=-5,bburx=45,bbury=5} 

		\hspace{-0.0cm}F
		\end{center}}
	\end{minipage}
	\hspace{2.4cm}\begin{minipage}[c]{0.5cm}
		\vbox{\begin{center}
		\epsfig{file=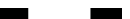,width=1.5cm,angle=270,
		bbllx=-10,bblly=-5,bburx=45,bbury=5} 

		\hspace{-0.0cm}G
		\end{center}}
	\end{minipage}
	\\
	\hline   
	$\sinh(\epsilon|B_x|/2)$
	&\hspace{1.4cm}\begin{minipage}[c]{1.5cm}
		\epsfig{file=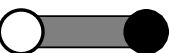,width=1.5cm,angle=270,
		bbllx=-5,bblly=0,bburx=55,bbury=85}
	\end{minipage}
	&\hspace{0.4cm}\begin{minipage}[c]{0.5cm}
		\vbox{\begin{center}
		\epsfig{file=table_4_3.eps,width=1.5cm,angle=270,
		bbllx=-10,bblly=-5,bburx=45,bbury=5} 

		\hspace{-0.0cm}G
		\end{center}}
	\end{minipage}
	\\
	\hline   
    \end{tabular}
\caption{\it Cluster break-ups of various plaquette and time-like bond 
configurations together with their relative probabilities $A,B,...,G$. Filled 
dots represent spin up, open circles represent spin down, and the fat lines are
the cluster connections.}
\end{center}
\end{table}

Eqs.(\ref{cluster1},\ref{cluster2}) can be viewed as a representation of the 
original model as a random cluster model. The cluster algorithm operates in two
steps. First, a cluster break-up is chosen for each space-time interaction 
plaquette or time-like interaction bond according to the above probabilities. 
This effectively replaces the original Boltzmann weight of a configuration with
a set of constraints represented by the $\delta$-functions associated with the 
chosen break-up. The constraints imply that the spins in one cluster can only
be flipped together. Second, every cluster is flipped with probability $1/2$. 
When a cluster is flipped, the spins on all sites that belong to the cluster 
are flipped from up to down and vice versa. Eqs.(\ref{balance1},\ref{balance2})
ensure that the cluster algorithm obeys detailed balance. To determine 
$A,B,...,E$ we distinguish three cases. For $J' > J > 0$ we solve 
eq.(\ref{balance1}) by
\begin{equation}
\label{solution1}
A = \exp(- \epsilon J'/2), \ B = \sinh(\epsilon J/2), \
C = \exp(- \epsilon J/2) - \exp(- \epsilon J'/2), \ D = E = 0.
\end{equation}
For $- J \leq J' \leq J$ we use
\begin{eqnarray}
\label{solution2}
&&A = \frac{1}{2}[\exp(- \epsilon J'/2) + \exp(- \epsilon J/2)],
\ B = \frac{1}{2}[\exp(\epsilon J/2) - \exp(- \epsilon J'/2)], \ 
C = 0, \nonumber \\
&&D = \frac{1}{2}[\exp(- \epsilon J'/2) - \exp(- \epsilon J/2)], \
E = 0, \end{eqnarray}
and for $J' \leq J < 0$
\begin{equation}
\label{solution3}
A = \cosh(\epsilon J/2), \ B = C = 0, \ D = \sinh(\epsilon |J|/2), \
E = \exp(- \epsilon J'/2) - \exp(\epsilon |J|/2).
\end{equation}
For example, for the antiferromagnetic quantum Heisenberg model ($J' = J > 0$)
we have
\begin{equation}
A = \exp(- \epsilon J/2), \, B  = \sinh(\epsilon J/2), \, C = D = E = 0.
\end{equation}
Consequently, on plaquette configurations $[\up,\up,\up,\up]$ or 
$[\down,\down,\down,\down]$ one always chooses time-like connections between 
the sites, and for configurations $[\up,\down,\down,\up]$ or 
$[\down,\up,\up,\down]$ one always chooses space-like connections. For 
configurations $[\up,\down,\up,\down]$ or $[\down,\up,\down,\up]$ one chooses 
time-like connections with probability $p = A/(A+B) = 2/[1 + \exp(\epsilon J)]$
and space-like connections with probability $1 - p = B/(A+B)$. Similarly, 
eq.(\ref{balance2}) yields
\begin{equation}
F = \exp(- \epsilon |B_x|/2), \, G = \sinh(\epsilon |B_x|/2).
\end{equation}

The above cluster rules were first used in a simulation of the Heisenberg 
antiferromagnet \cite{Wie94} in the absence of a magnetic field. In that case
there is no sign problem. Then the corresponding loop-cluster algorithm is 
extremely efficient and has almost no detectable autocorrelations. When a 
magnetic field is switched on the situation changes. When the magnetic field 
points in the direction of the spin quantization axis (the $3$-direction in our
case) there is no sign problem. However, the magnetic field then explicitly 
breaks the $\Z(2)$ flip symmetry on which the cluster algorithm is based, and 
the clusters can no longer be flipped with probability $1/2$. Instead the flip 
probability is determined by the value of the magnetic field and by the 
magnetization of the cluster. When the field is strong, flips of magnetized
clusters are rarely possible and the algorithm becomes very inefficient. To 
avoid this problem, we have chosen the magnetic field to point in the 
$1$-direction, i.e. perpendicular to the spin quantization axis. In that case, 
the cluster flip symmetry is not affected by the magnetic field, and the 
clusters can still be flipped with probability $1/2$. However, in several cases
one now has a sign problem and the cluster algorithm becomes extremely 
inefficient again. Fortunately, using the meron concept the sign problem can be
eliminated completely and the efficiency of the original cluster algorithm can 
be maintained even in the presence of a magnetic field.

\section{Meron-Clusters and the Sign Problem}

Let us consider the effect of a cluster flip on the sign. As discussed before,
the flip of a meron-cluster changes $\Sign[s]$, while the flip of a 
non-meron-cluster leaves $\Sign[s]$ unchanged. An example of a meron-cluster is
shown in figure 2 along with the same spin configurations as in figure 1. When 
the meron cluster is flipped the first configuration with $\Sign[s] = 1$ turns 
into the second configuration with $\Sign[s] = - 1$. This property of the 
cluster is independent of the orientation of any other cluster. Since flipping 
all spins leaves $\Sign[s]$ unchanged, the total number of meron-clusters is 
always even.
\begin{figure}[htb]
\hbox{
\hspace{2.7cm}
${\rm Sign[s]}=1$
\hspace{5.0cm}
${\rm Sign[s]}=-1$
}  
\begin{center}
\hbox{
\epsfig{file=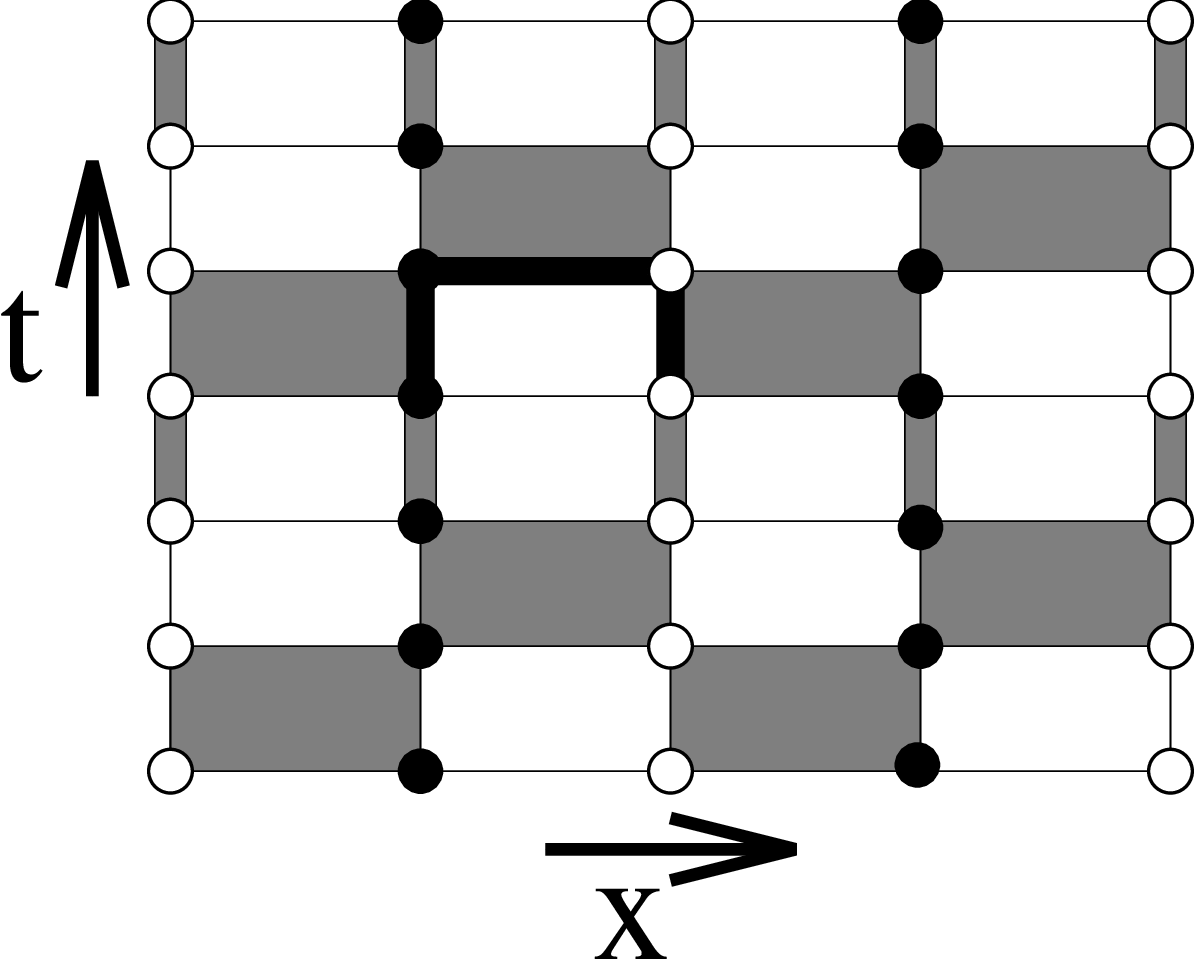,
width=4.5cm,angle=270,
bbllx=0,bblly=0,bburx=225,bbury=337}
\hspace{1.0cm}
\epsfig{file=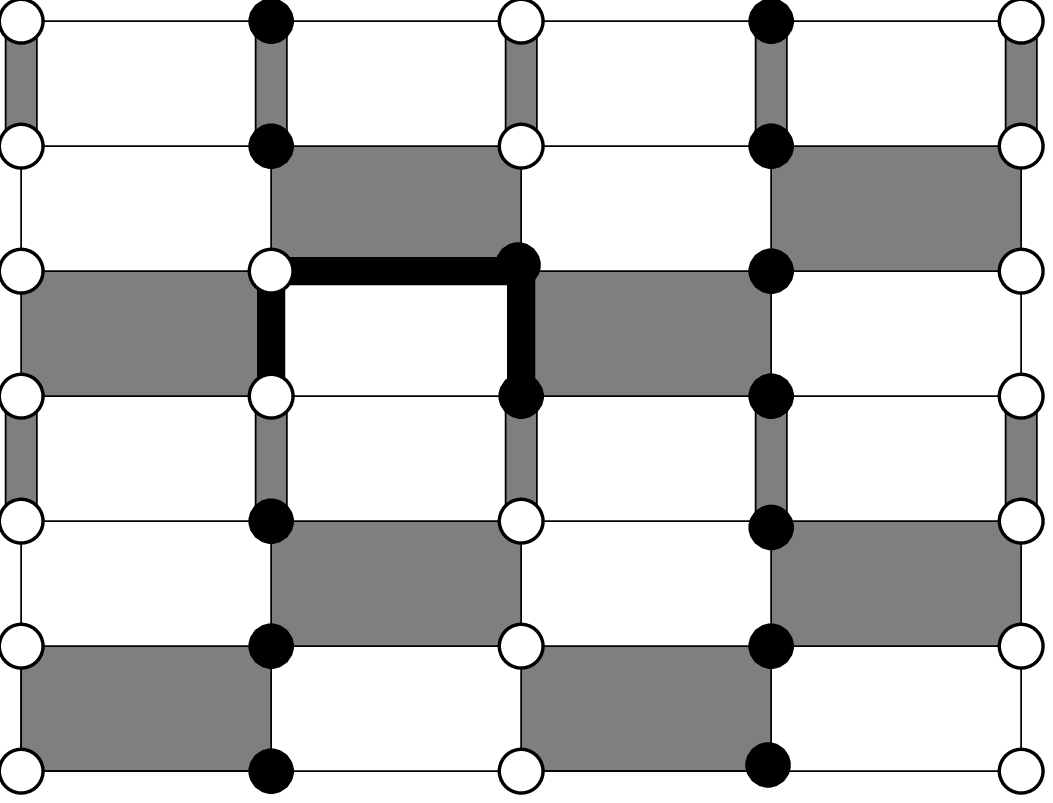,
width=4.5cm,angle=270,
bbllx=0,bblly=0,bburx=225,bbury=337}
}  
\end{center}
\caption{\it The same spin configurations as in figure 1 together with a 
meron-cluster represented by the fat black line. The other clusters are not 
shown. Flipping the meron-cluster changes one configuration into the other and 
changes $\Sign[s]$.}
\end{figure}

The meron concept allows us to gain an exponential factor in statistics. Since 
all clusters can be flipped independently with probability $1/2$, one can 
construct an improved estimator for $\langle \Sign \rangle$ by averaging 
analytically over the $2^{N_C}$ configurations obtained by flipping the $N_C$ 
clusters in a configuration in all possible ways. For configurations that 
contain merons, the average $\Sign[s]$ is zero because flipping a single 
meron-cluster leads to a cancellation of contributions $\pm 1$. Hence only the
configurations without merons contribute to $\langle \Sign \rangle$. The 
probability to have a configuration without merons is equal to 
$\langle \Sign \rangle$ and is hence exponentially suppressed with the 
space-time volume. The vast majority of configurations contains merons and 
contributes an exact $0$ to $\langle \Sign \rangle$ instead of a statistical 
average of contributions $\pm 1$. In this way the improved estimator leads to 
an exponential gain in statistics.

Now we want to show that the contributions from the zero-meron sector are 
always positive. With no merons in the configuration it is clear that the sign 
remains unchanged under cluster flips, but it is not obvious that one always 
has $\Sign[s] = 1$. To prove this we note that there is no sign problem in the
absence of the magnetic field. In that case all configurations have 
$\Sign[s] = 1$ and all clusters are closed loops. In the presence of the 
magnetic field some of the closed loops are cut into open string pieces. 
However, one can always flip the individual pieces such that a typical 
configuration of the model without the magnetic field emerges. These reference
configurations have $\Sign[s] = 1$. Since any configuration can be turned into 
such a configuration by cluster flips, and since the sign remains unchanged 
when a non-meron-cluster is flipped, all configurations without merons have 
$\Sign[s] = 1$. At this point we have solved one half of the sign problem. 

Before we can solve the other half of the problem we must discuss the improved 
estimator for the magnetization. Since the magnetization operator 
$M^1 = \sum_x S^1_x$ is not diagonal in the basis we have chosen to quantize 
the spins (which is along the $3$-direction), it is not obvious how to measure 
it. However, as discussed in \cite{Bro98}, quantum cluster algorithms naturally
allow us to measure non-diagonal operators. In the present case, due to time 
translation invariance, $M^1$ is equivalent to a sum of local spin-flip
operators $S_x^1$ inserted at any time slice and appropriately averaged. An 
open string cluster that contains the point $(x,t)$ is naturally divided into 
two parts, one on each side of the point. Flipping each side while keeping the 
other side fixed produces spin configurations that contribute to the spin flip 
operator $S_x^1$ at time $t$. Closed loop-clusters, on the other hand, do not 
decompose into two parts and thus do not contribute to this operator. If we 
construct an improved estimator for $S_x^1$ inserted at the time $t$, a 
configuration containing more than one meron does not contribute to 
$\langle S_x^1\rangle$. Since the total number of merons is always even, in 
there exists at least one meron that does not contain the point $(x,t)$. 
Flipping that meron changes the sign but not $S_x^1$ and hence results in an 
exact cancellation of two equal and opposite contributions to 
$\langle S_x^1\rangle$. Hence we can focus entirely on the zero-meron sector. 
In this sector the point $(x,t)$ can divide an open string cluster into two 
merons or two non-merons. When both parts are non-merons, flipping one part but
not the other yields a contribution of $1/2$ to $\langle S_x^1 \rangle$. When 
both parts are merons, on the other hand, one gets a contribution of $-1/2$. 
Summing over $x,t$ one obtains $\langle M^1 \rangle$. 

For an antiferromagnet in a uniform magnetic field $B_x = B$, one can find a 
simple expression for the improved estimator for $\langle M^1 \rangle$ in terms
of a winding number of closed loops which result from joining the open string 
clusters. In that case a non-meron cluster has the property that both end 
points of an open string cluster have the same spin type, either up or down. If
all the open string clusters are flipped into a reference configuration and 
joined together to form closed loops, one can define $W_l$ for each loop to be 
the temporal winding number of the loop. We assume that the cluster is growing 
in the positive direction on the sites where the loop is broken into open 
strings. Since the loop is always broken up on the same type of spin, the above
definition of the winding of a given loop is consistent. Further, if a 
particular loop is not composed of open string clusters then $W_l = 0$. With 
this definition of $W_l$ it is easy to show that
\begin{equation}
\label{ratio}
\langle M^1 \rangle = 
\frac{\langle \delta_{N,0} \sum_l W_l \rangle}{2 \langle \delta_{N,0} \rangle},
\end{equation}
where $N$ is the number of meron-clusters. Hence, as explained above, the
magnetization gets non-vanishing contributions only from the zero-meron sector.
One should note that $W_l$ can be negative.

Since the magnetization gets non-vanishing contributions only from the 
zero-meron sector, it is unnecessary to generate any configuration that 
contains meron-clusters. This observation is the key to the solution of the 
other half of the sign problem. In fact, one can gain an exponential factor in 
statistics by restricting the simulation to the zero-meron sector, which 
represents an exponentially small fraction of the whole configuration space. 
This procedure enhances both the numerator and the denominator in 
eq.(\ref{ratio}) by a factor that is exponentially large in the volume, but 
leaves the ratio of the two invariant. In practice, it is advantageous to 
occasionally generate configurations containing merons even though they do not 
contribute to our observable, because this reduces the autocorrelation times. 
Still, in order to solve the sign problem multi-meron configurations must be 
very much suppressed. To suppress such configurations in a controlled way, we 
introduce an additional Boltzmann factor $q < 1$ for each meron, which 
determines the relative weights of the $N$-meron sectors. We visit all 
plaquette and time-like bond interactions one after the other and choose new 
pair connections between the sites according to the above cluster rules. The 
suppression factor $q$ is used in a Metropolis accept-reject step. A newly 
proposed pair connection that changes the number of merons from $N$ to $N'$ is 
accepted with probability $p = \mbox{min}[1,q^{N'-N}]$. It should be stressed 
that adding the Boltzmann factor for the reweighting doesn't change the 
physics, only the algorithm. After visiting all plaquette and time-like bond 
interactions, each cluster is flipped with probability $1/2$ which completes 
one update sweep. After reweighting, multi-meron configurations are very much 
suppressed. This completes the second step in our solution of the sign problem.

\section{Antiferromagnetic Spin Ladders in a Uniform Magnetic Field} 

Antiferromagnetic spin ladders --- sets of several transversely coupled quantum
spin chains --- are interesting condensed matter systems which interpolate 
between single 1-d spin chains and 2-d quantum antiferromagnets. The ladders 
are spatially quasi 1-d systems whose low-energy dynamics are governed by 
$(1+1)$-d quantum field theories, which can be solved analytically using Bethe 
ansatz techniques. This allows us to test field theoretical predictions for 
such models using condensed matter experiments as well as numerical 
simulations. Here we consider spin ladders in an external uniform magnetic 
field $B_x = B$, which corresponds to a chemical potential $\mu = B/c$ in the 
corresponding $(1+1)$-d quantum field theory. The Bethe ansatz solution of this
theory yields predictions for the magnetization of the ladder system, which we 
compare with results of numerical simulations.

Haldane was first to realize the connection between spin chains and $(1+1)$-d
field theories \cite{Hal83}. He conjectured that a single 1-d antiferromagnetic
quantum spin chain of length $L$ at inverse temperature $\beta$ is described by
a $(1+1)$-d $O(3)$ symmetric quantum field theory with an action
\begin{equation}
S[\vec e] = \int_0^\beta dt \int_0^L dx \
[\frac{c}{2g^2}(\partial_x \vec e \cdot \partial_x \vec e 
+ \frac{1}{c^2} \partial_t \vec e \cdot \partial_t \vec e)
+ \frac{i \theta}{4 \pi} \vec e \cdot 
(\partial_x \vec e \times \partial_t \vec e)].
\end{equation}
Here $\vec e(x,t)$ is a three-component unit-vector field, $g$ is a coupling 
constant and $c$ is the spin-wave velocity. The vacuum angle is given by 
$\theta = 2 \pi S$, where $S$ is the value of the spin, hence quantum spin 
chains with integer spin have $\theta = 0$ and chains with half-integer spin 
have $\theta = \pi$. The $(1+1)$-d $O(3)$ model is an asymptotically free 
quantum field theory with a non-perturbatively generated mass-gap 
$m \propto \exp(- 2 \pi/g^2)$ at $\theta = 0$. At $\theta = \pi$, on the other 
hand, the mass-gap disappears. This naturally explains why chains of 
half-integer spins are gapless, while chains of integer spins have a gap. For 
large values of the spin $S$ the coupling constant is given by $1/g^2 = S/2$, 
so the mass-gap $m \propto \exp(- \pi S)$ goes to zero and the system 
approaches a continuum limit.
 
Chakravarty, Halperin and Nelson used a $(2+1)$-d effective field theory to
describe the low-energy dynamics of spatially 2-d quantum antiferromagnets
\cite{Cha88}. Chakravarty has applied this theory to quantum spin ladders with 
a large even number of coupled spin $1/2$ chains \cite{Cha96}. We consider spin
ladders with the same value of the antiferromagnetic coupling along and between
the chains. These systems are described by the action
\begin{equation}
S[\vec e] = \int_0^\beta dt \int_0^L dx \int_0^{L'} dy \
\frac{\rho_s}{2}[\partial_x \vec e \cdot \partial_x \vec e 
+ \partial_y \vec e \cdot \partial_y \vec e
+ \frac{1}{c^2} \partial_t \vec e \cdot \partial_t \vec e],
\end{equation}
where $\rho_s$ is the spin stiffness and $c$ is the spin-wave velocity. The 
coupled spin chains are oriented in the spatial $x$-direction with a large 
extent $L$, while the transverse $y$-direction has a much smaller extent 
$L' \ll L$. Here we consider spin ladders with periodic boundary conditions in 
the transverse direction, and we limit ourselves to an even number of coupled 
spin $1/2$ chains. The effective action for a ladder with an odd number of 
coupled chains would contain an additional topological term.

When the spin ladder is placed in a uniform external magnetic field $\vec B$,
the field couples to a conserved quantity --- the uniform magnetization. Hence,
on the level of the effective theory, the magnetic field plays the role of a
chemical potential, i.e. it appears as the time-component of an imaginary
constant vector potential. As a consequence, the ordinary derivative
$\partial_t \vec e$ is replaced by the covariant derivative $\partial_t \vec e
+ i \vec B \times \vec e$ and the effective action takes the form
\begin{equation}
S[\vec e] = \int_0^\beta dt \int_0^L dx \int_0^{L'} dy \ 
\frac{\rho_s}{2}[\partial_x \vec e \cdot \partial_x \vec e 
+ \partial_y \vec e \cdot \partial_y \vec e
+ \frac{1}{c^2} (\partial_t \vec e + i \vec B \times \vec e) \cdot
(\partial_t \vec e + i \vec B \times \vec e)].
\end{equation}

For a sufficiently large even number of coupled chains ($L' \gg c/\rho_s$) the 
ladder system undergoes dimensional reduction to the $(1+1)$-d $O(3)$ symmetric
quantum field theory with the action
\begin{eqnarray}
S[\vec e]&=&\int_0^\beta dt \int_0^L dx \
\frac{\rho_s L'}{2}[\partial_x \vec e \cdot \partial_x \vec e 
+ \frac{1}{c^2} (\partial_t \vec e + i \vec B \times \vec e) \cdot
(\partial_t \vec e + i \vec B \times \vec e)] \nonumber \\
&=&\int_0^{\beta c} dct \int_0^L dx \
\frac{1}{2g^2}[\partial_x \vec e \cdot \partial_x \vec e 
+ (\partial_{ct} \vec e + i \vec \mu \times \vec e) \cdot
(\partial_{ct} \vec e + i \vec \mu \times \vec e)].
\end{eqnarray}
The effective coupling constant is given by $1/g^2 = \rho_s L'/c$ and the 
magnetic field appears as a chemical potential of magnitude $\mu = B/c$.

The $(1+1)$-d $O(3)$ model with chemical potential $\mu$ has been solved 
exactly by Wiegmann \cite{Wie85} using Bethe ansatz techniques. Using this
solution, Hasenfratz, Maggiore and Niedermayer have derived an exact result for
the non-perturbatively generated mass-gap $m$ in this model \cite{Has90}. In 
particular, for $\mu \geq m$ and $\beta = L = \infty$ the Bethe ansatz yields 
the free energy density
\begin{equation}
\label{free}
f(\mu) - f(0) = 
- \frac{1}{2 \pi} \int_{-\theta_0}^{\theta_0} d\theta \epsilon(\theta) m
\cosh\theta,
\end{equation}
where the function $\epsilon(\theta)$ satisfies the integral equation
\begin{equation}
\label{epsilon}
\epsilon(\theta) - \int_{-\theta_0}^{\theta_0} d\theta' 
\frac{\epsilon(\theta')}{(\theta' - \theta)^2 + \pi^2} = \mu - m \cosh\theta,
\end{equation}
and $\theta_0$ is defined by the boundary condition 
$\epsilon(\pm \theta_0) = 0$. For $\mu < m$ no particles can be produced and 
$f(\mu) = f(0)$. For chemical potentials slightly above this threshold the 
previous equations imply
\begin{equation}
f(\mu) - f(0) = - \frac{2}{3} \frac{\sqrt{2m}}{\pi} (\mu - m)^{3/2}
\end{equation}
and the resulting particle density is given by
\begin{equation}
\frac{\langle N \rangle}{L} = - \frac{df(\mu)}{d\mu} = \frac{\sqrt{2m}}{\pi} 
\sqrt{\mu - m}.
\end{equation}
This quantum field theoretic result can be translated back into the language of
condensed matter physics. It yields an expression for the magnetization density
\begin{equation}
\label{root}
\frac{\langle M^1 \rangle}{L} = \frac{\sqrt{2m}}{\pi} \sqrt{\frac{B}{c} - m}
\end{equation}
as a function of the magnetic field $B$. The square root form was derived 
earlier in \cite{Chi97}. Our result yields an explicit expression for the 
prefactor as well. 

In the next chapter we will present results of numerical simulations using the
meron-cluster algorithm. The numerical results are obtained at small but
non-zero temperatures and for ladders of large but finite lengths, while the
above analytic expressions are for $\beta = L = \infty$. In particular, the 
results in the threshold region $\mu \approx m$ are very sensitive to 
finite-size and finite-temperature effects. Fortunately, these effects can 
still be understood analytically. In particular, the Bethe ansatz solution 
reveals that for not too large $\mu$ the system is equivalent to a dilute gas 
of fermions. Its magnetization density is hence given by
\begin{equation}
\label{finitesize}
\frac{\langle M^1 \rangle}{L} = \frac{1}{L} \sum_{n \in \Z}
[1 + \exp(\beta (c \sqrt{(\frac{2 \pi n}{L})^2 + m^2} - B))]^{-1}.
\end{equation}

Up to this point we have considered the mass-gap $m$ as an unknown constant. 
Based on results from \cite{Cha88,Has90,Has91}, Chakravarty \cite{Cha96} has 
expressed the mass-gap of the ladder system as
\begin{equation}
m = \frac{16 \pi \rho_s}{e c} \exp(- 2 \pi \rho_s L'/c)
[1 + \frac{c}{4 \pi \rho_s L'} + {\cal O}((\frac{c}{\rho_s L'})^2)].
\end{equation}
For the spin $1/2$ quantum Heisenberg model with exchange coupling $J$ on a 
2-d lattice of spacing $a$, the values of the spin stiffness 
$\rho_s = 0.1800(5) J$ and the spin-wave velocity $c = 1.657(2) Ja$ are known 
from very precise cluster algorithm simulations \cite{Bea98}. As discussed in 
\cite{Bea98}, the above expression for the mass-gap is expected to be accurate 
for a sufficiently large number $L'/a > 12$ of coupled chains. The simulations
in the present paper are performed at $L'/a = 4$ and are thus not expected to
be described accurately by this formula. Still, for $L'/a = 4$ the value 
$m = 0.141(2)/a$ is known from the simulations of \cite{Syl97}.

There is another interesting phenomenon that occurs at rather large values of
the magnetic field, namely saturation of the magnetization --- all spins follow
the external field and the system becomes completely ferromagnetically ordered.
This effect cannot be understood in the framework of the above low-energy 
effective theory because it assumes antiferromagnetic order. Still, it is easy 
to convince oneself that there must be a critical magnetic field beyond which 
saturation occurs. This effect is indeed observed in the numerical simulations 
described below.

\section{Numerical Results}

\begin{figure}[htb]
\begin{minipage}{\textwidth}
\begin{center}
\epsfig{file=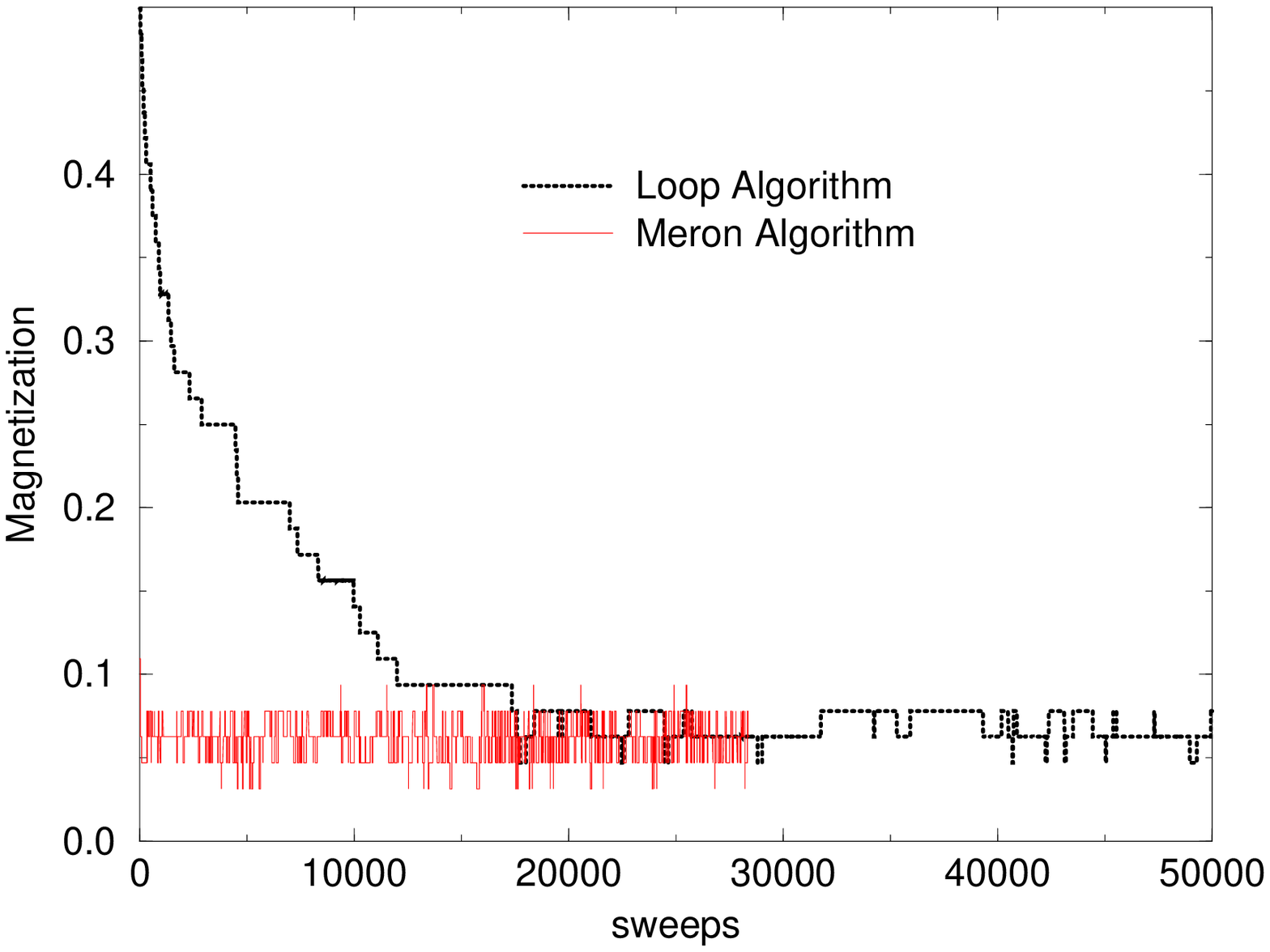,
width=9cm,angle=0,
bbllx=12,bblly=12,bburx=577,bbury=448}

\epsfig{file=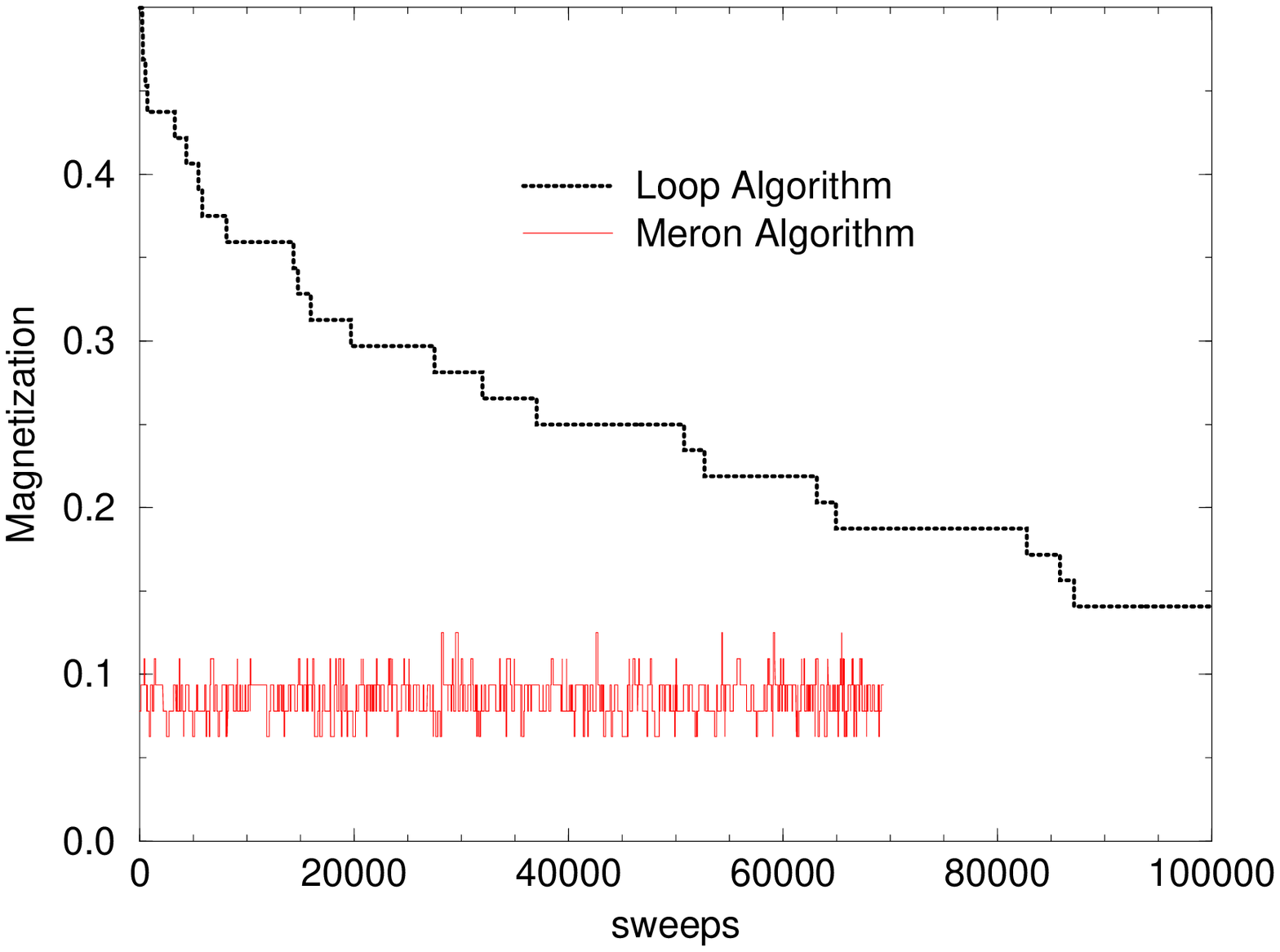,
width=9cm,angle=0,
bbllx=12,bblly=12,bburx=577,bbury=448}
\end{center}
\end{minipage}
\caption{\it Thermalization of the magnetization for the loop-cluster algorithm
versus the meron-cluster algorithm. The first graph is for a magnetic field 
$B = 0.75J$ and the second graph is for $B = J$. For $B = J$ the loop algorithm
takes more than 100000 sweeps to reach equilibrium while the meron-cluster 
algorithm has no thermalization problem.}
\end{figure}
We have performed numerical simulations with the meron-cluster algorithm for
various quantum antiferromagnets in a uniform magnetic field $B$. To 
demonstrate the efficiency of the algorithm, we have compared it with the
standard loop-cluster algorithm. In case of the loop algorithm, the magnetic
field points in the direction of the spin quantization axis. Hence the $\Z(2)$
symmetry that allows the clusters to be flipped with probability $1/2$ at $B=0$
is explicitly broken by the magnetic field. In this case some loop-clusters can
be flipped only with very small probability and the algorithm becomes
inefficient for large values of $B$. In the meron-cluster algorithm, on the
other hand, the magnetic field is perpendicular to the spin quantization axis
and all clusters can still be flipped with probability $1/2$. Now the sign 
problem arises, but it is solved completely by the meron-cluster algorithm. 
Figure 3 compares the thermalization behavior of the magnetization of a 2-d 
Heisenberg antiferromagnet on an $8 \times 8$ lattice at $\beta J = 10$ with 
$M = 100$ for the loop-cluster algorithm and the meron-cluster algorithm. At 
$B=0.75J$, the loop-cluster algorithm takes about 20000 sweeps until thermal 
equilibrium is reached, while the meron-cluster algorithm equilibrates much
faster. At $B=J$, the loop-cluster algorithm already needs more than 100000
equilibration sweeps, but the meron-cluster algorithm again has no 
thermalization problem. It should be pointed out that the loop-cluster 
algorithm works reasonably well for small values of $B$ and has been used in an
interesting numerical study in \cite{Tro98}. However, for larger values of $B$ 
the meron-cluster algorithm is clearly superior. It would be interesting to
compare the performance of the meron-cluster algorithm with that of the worm
algorithm \cite{Pro96,Kas99} and the stochastic series expansion technique 
\cite{San99}.

Some of our data for the magnetization density $\langle M^1 \rangle a/L$ of 
2-d antiferromagnetic spin systems on an $L/a \times L'/a$ lattice at inverse 
temperature $\beta$ are contained in table 2.
\begin{table}
\begin{center}
\begin{tabular}{|c|c|c|c|c|c|}
\hline
$L/a$ & $L'/a$ & $\beta J$ & $M$ & $B/J$ & $\langle M^1 \rangle a/L$ \\
\hline
\hline
  8 & 8 & 10 & 100 & 0.75 & 0.488(6) \\
\hline
  8 & 8 & 10 & 100 & 1.00 & 0.690(6) \\
\hline
  20 & 4 & 15 & 200 & 0.10 & 0.0048(4) \\
\hline
  20 & 4 & 15 & 200 & 0.20 & 0.0184(4) \\
\hline
  20 & 4 & 15 & 200 & 0.30 & 0.0452(8) \\
\hline
  20 & 4 & 15 & 200 & 0.40 & 0.086(2) \\
\hline
  20 & 4 & 15 & 200 & 0.50 & 0.120(4) \\
\hline
  20 & 4 & 15 & 200 & 1.00 & 0.324(4) \\
\hline
  20 & 4 & 15 & 200 & 2.00 & 0.76(2) \\
\hline
  20 & 4 & 15 & 200 & 3.00 & 1.280(8) \\
\hline
  20 & 4 & 15 & 200 & 4.00 & 1.93(3) \\
\hline
  20 & 4 & 15 & 200 & 4.20 & 2.000(8) \\
\hline
  20 & 4 & 15 & 200 & 4.40 & 2.000(8) \\
\hline
  40 & 4 & 24 & 300 & 0.10 & 0.00104(8) \\
\hline
  40 & 4 & 24 & 300 & 0.20 & 0.0096(6) \\
\hline
  40 & 4 & 24 & 300 & 0.30 & 0.042(2) \\
\hline
  40 & 4 & 24 & 300 & 0.40 & 0.085(3) \\
\hline
  40 & 4 & 24 & 300 & 0.50 & 0.117(7) \\
\hline
  40 & 4 & 24 & 300 & 1.00 & 0.332(4) \\
\hline
\end{tabular}
\end{center}
\caption{\it Numerical results for the magnetization density 
$\langle M^1 \rangle a/L$ for various spatial sizes $L,L'$, inverse 
temperatures $\beta$, Trotter numbers $M$ and magnetic field values $B$.}
\end{table}
\begin{figure}[htb]
\vbox{
\begin{center}
\psfig{figure=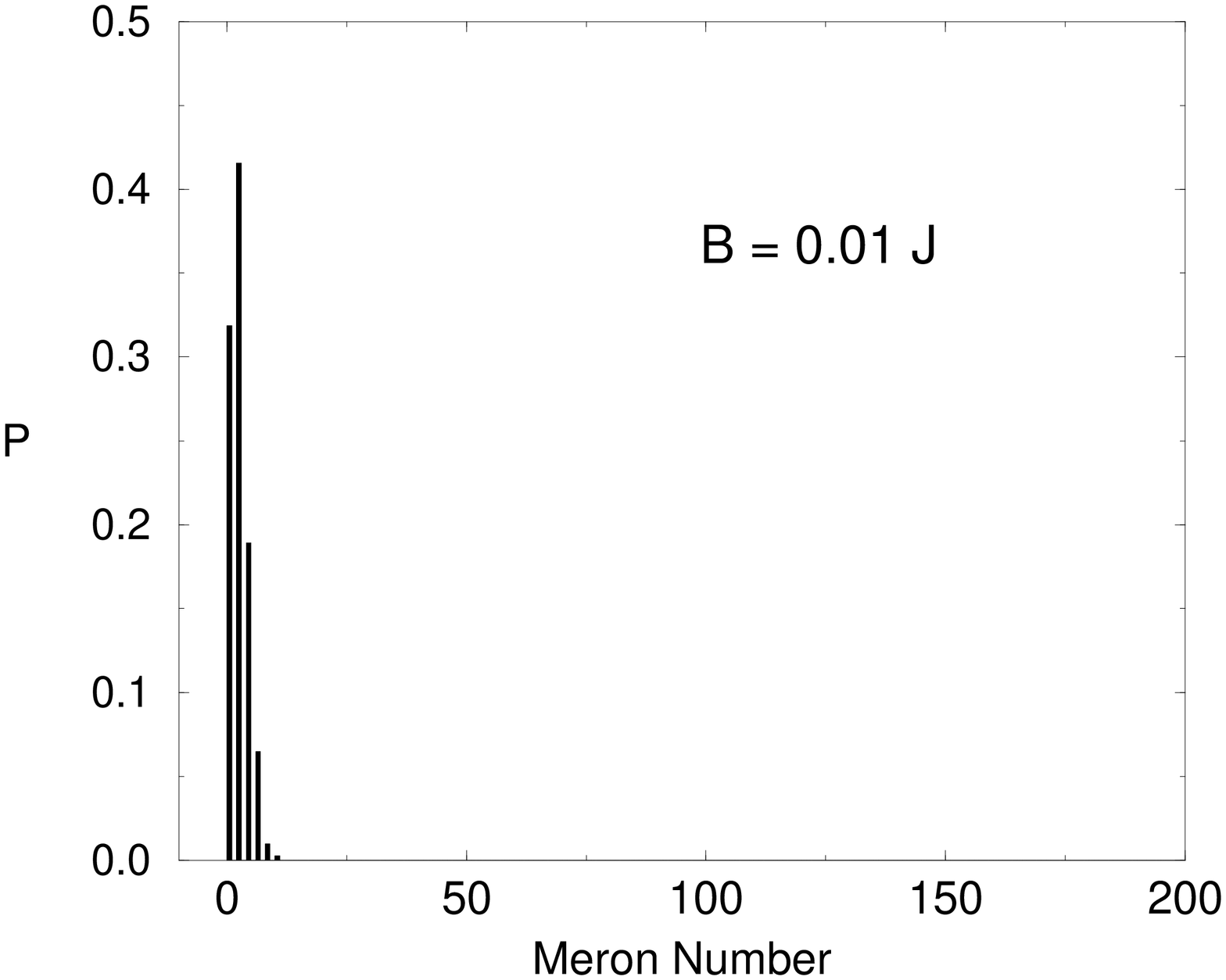,height=2.5in,width=2in,angle=270}
\psfig{figure=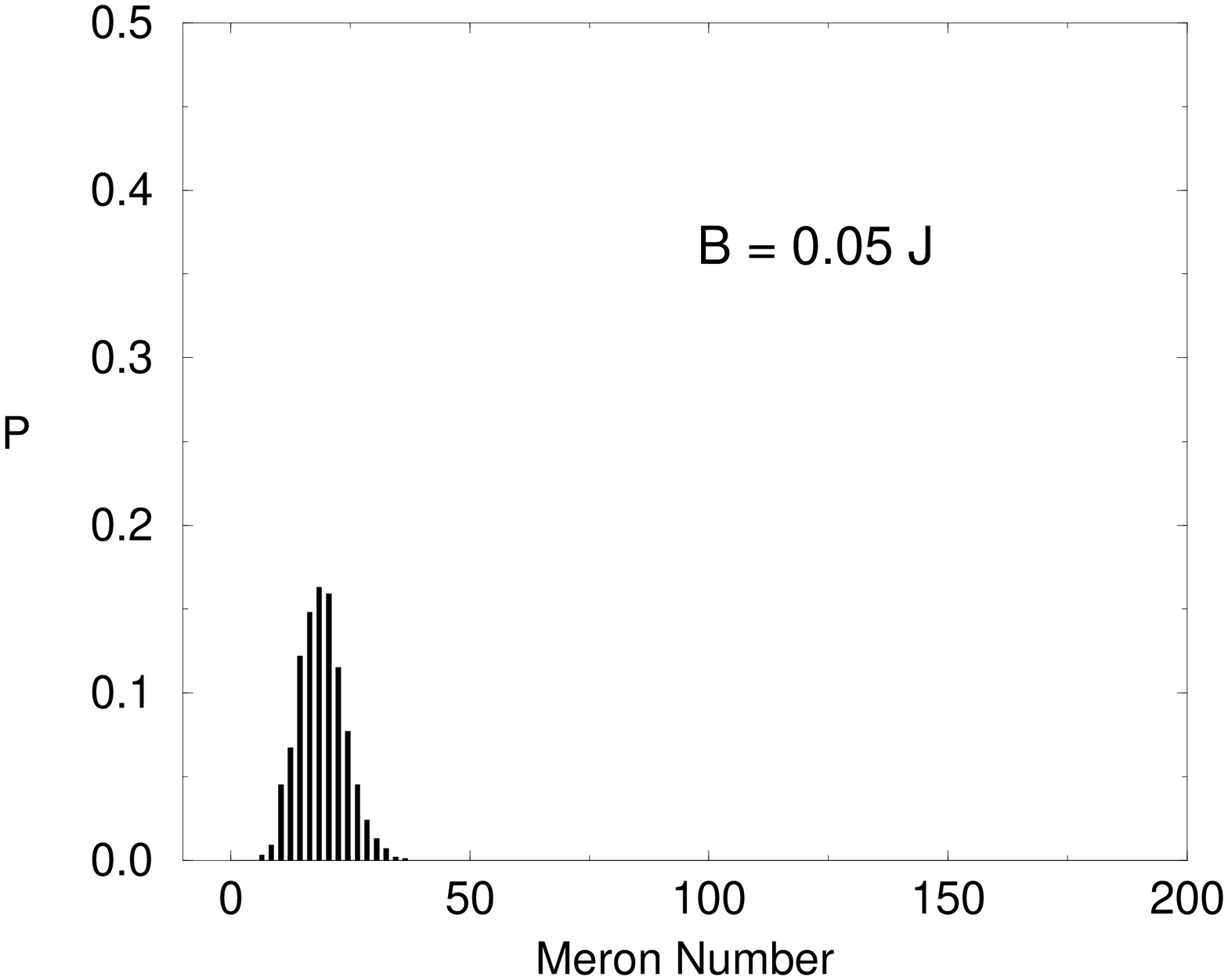,height=2.5in,width=2in,angle=270}
\psfig{figure=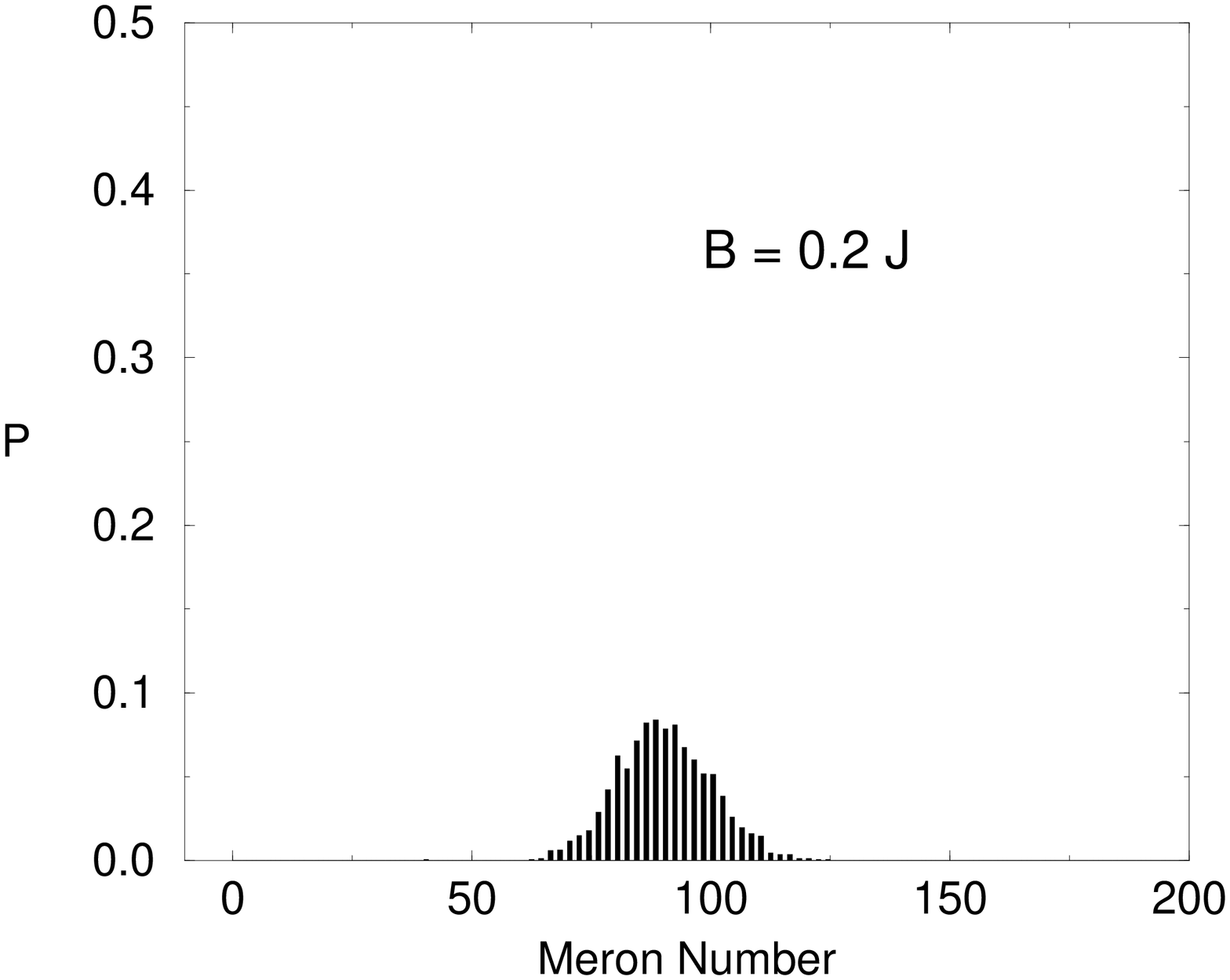,height=2.5in,width=2in,angle=270}
\psfig{figure=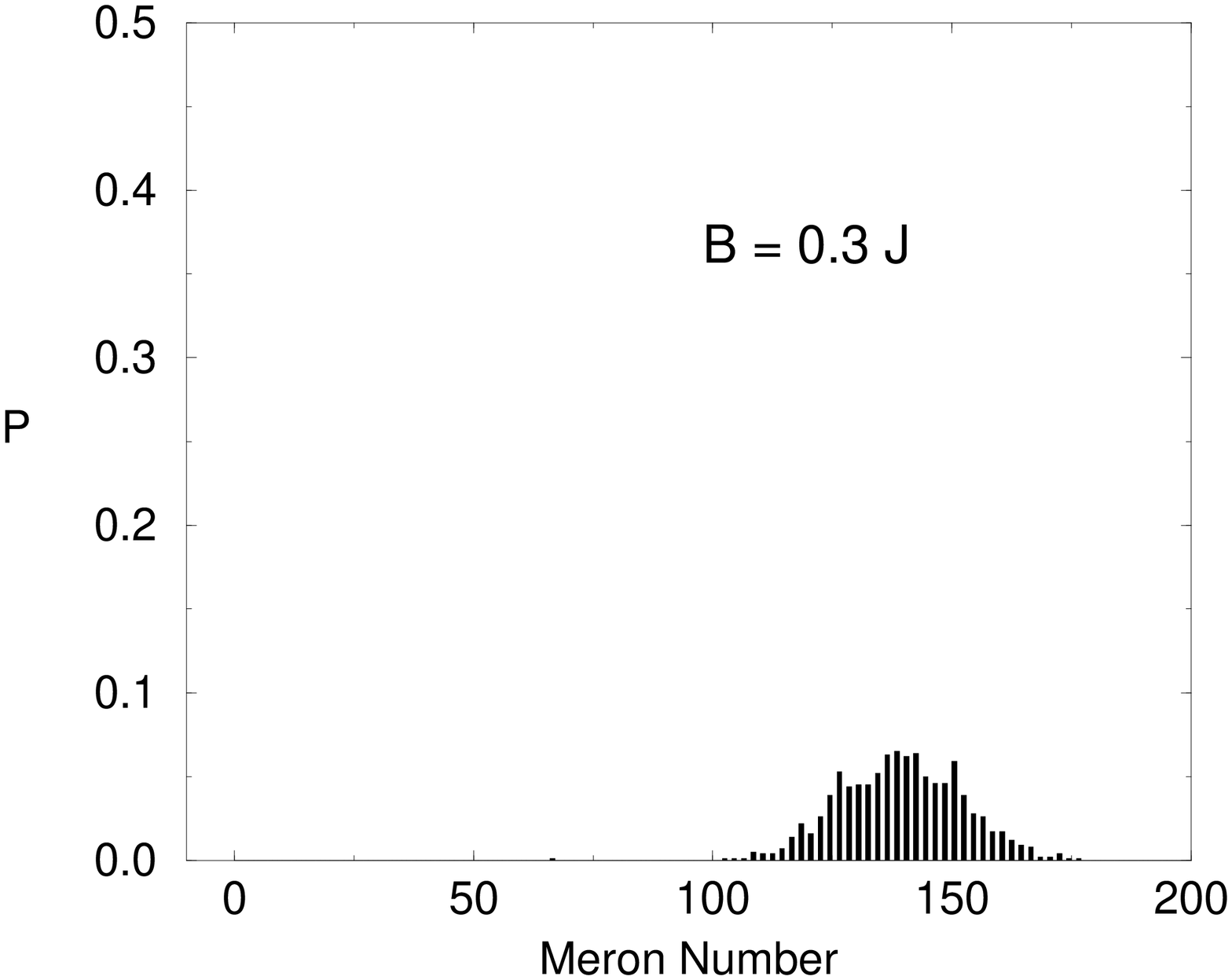,height=2.5in,width=2in,angle=270}
\caption{\it The probability of having a certain number of merons for various
values of the magnetic field $B/J = 0.01, 0.05, 0.2$ and $0.3$.}
\end{center}
}
\end{figure}
Figure 4 shows the probability to have a certain number of merons in an 
algorithm that samples all meron-sectors without reweighting. This calculation
was done on an $8 \times 8$ lattice at $\beta J = 1$ and $M = 100$. For small 
values of $B$ the zero-meron sector and hence $\langle \Sign \rangle$ are 
relatively large, while multi-meron configurations are rare. On the other hand,
for larger values of the magnetic field the vast majority of configurations
have a large number of merons and hence $\langle \Sign \rangle$ is 
exponentially small. The meron-cluster algorithm suppresses the multi-meron 
configurations and concentrates on the zero-meron sector which is the only one 
that contributes to the magnetization.

\begin{figure}[htb]
\begin{center}
\epsfig{file=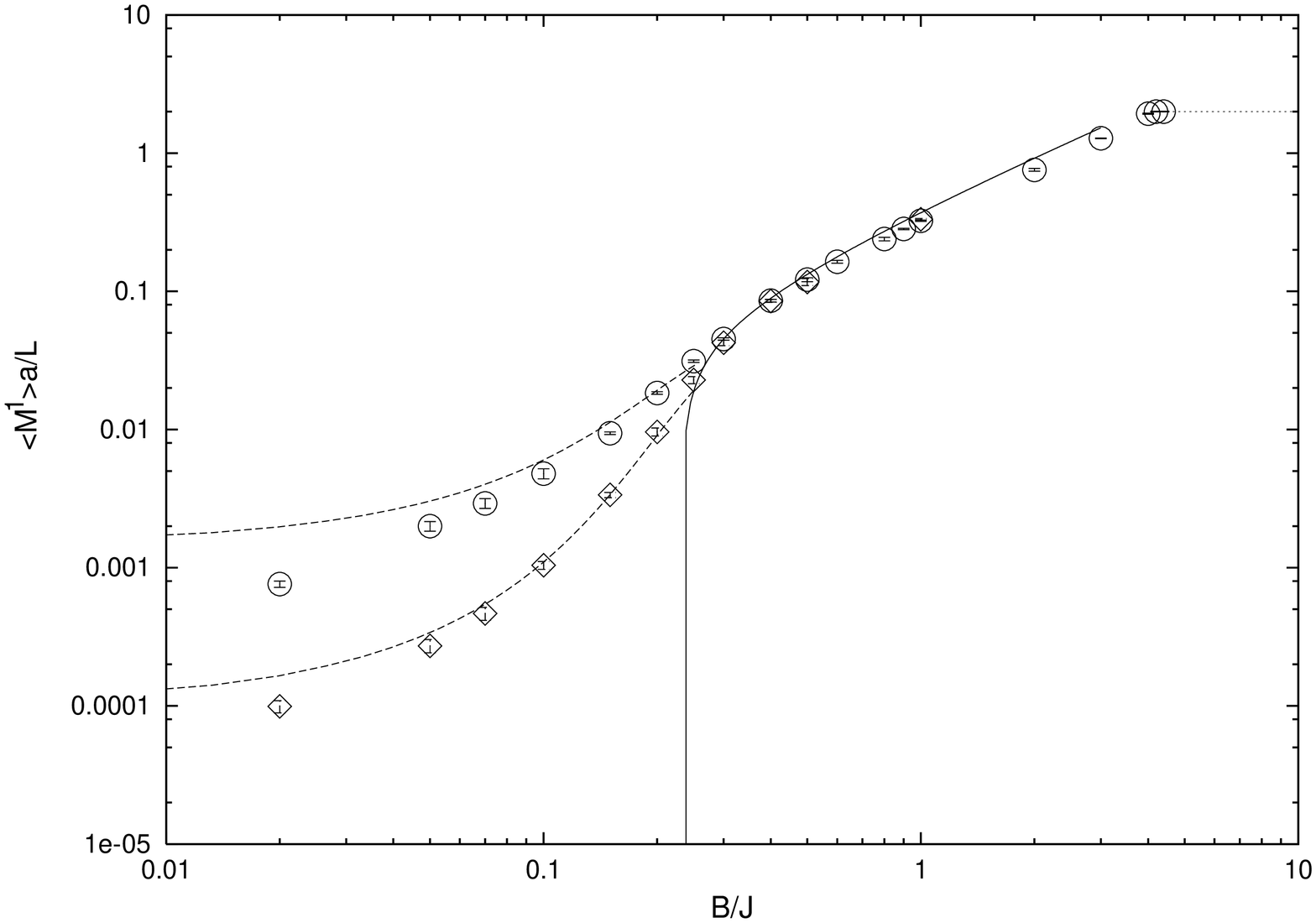,
width=8.5cm,angle=270,
bbllx=53,bblly=50,bburx=554,bbury=770}
\end{center}
\caption{\it Magnetization density $\langle M^1 \rangle/L$ of quantum spin 
ladders consisting of $L'/a = 4$ coupled chains as a function of the magnetic 
field $B$. The numerical data are for two systems: one of size $L/a = 20$ at 
inverse temperature $\beta J = 15$ (circles) and the other for $L/a = 40$ at 
$\beta J = 24$ (diamonds). The solid curve is the infinite volume, zero 
temperature analytic result, while the two dashed curves are finite volume, 
non-zero temperature analytic results for the two simulated systems in the 
small $B$ region. The dotted curve represents saturation of the magnetization
per spin at $1/2$.}
\end{figure}
Figure 5 shows the magnetization density of antiferromagnetic quantum spin 
ladders consisting of four coupled spin $1/2$ chains (i.e. $L'/a = 4$). The
cases $L/a = 20, \beta J = 15, M = 200$ and $L/a = 40, \beta J = 24, M = 300$ 
have been investigated. Within our statistical errors these Trotter numbers
give results indistinguishable from the time-continuum limit. In any case, 
following \cite{Bea96} it would be straightforward to implement the
meron-cluster algorithm directly in the Euclidean time continuum. Using the 
values $m = 0.141(2)/a$ from \cite{Syl97} and $c = 1.657(2) Ja$ from 
\cite{Bea98}, the finite volume, non-zero temperature expression of 
eq.(\ref{finitesize}) that was derived for small values of $B$ (represented by
the dashed curves in figure 5) describes the data rather well in that regime. 
Using the same values for $m$ and $c$, the infinite volume, zero temperature 
expression for $\langle M^1 \rangle/L$ that results from solving 
eqs.(\ref{free}) and (\ref{epsilon}) (represented by the solid curve in figure
5) is in good agreement with the numerical data for intermediate values of $B$.
For $B/c$ slightly above the threshold $m$, the infinite volume, zero 
temperature expression for $\langle M^1 \rangle/L$ is given by eq.(\ref{root}).
In the threshold region finite size and finite temperature effects are rather 
large and our numerical data do not fall in the applicability range of 
eq.(\ref{root}). As expected, for large values of $B$ the magnetization per 
spin saturates at $1/2$ (represented by the dotted curve in figure 5). Note 
that $\langle M^1 \rangle a/L = 2$ for $L'/a = 4$ in table 2 corresponds to a 
value $\langle M^1 \rangle a^2/(L L') = 1/2$ per spin. It should be stressed 
that the comparison of analytic and numerical results does not involve any 
adjustable parameters. Of course, one should not expect perfect agreement. 
Given the fact that the simulations were performed at $L'/a = 4$ while the 
analytic expressions were derived in the large $L'/a$ limit, the agreement 
between theory and numerical data is quite remarkable.

\section{Conclusions}

Antiferromagnetic quantum spin systems in a uniform external magnetic field $B$
are very difficult to simulate numerically. For example, the standard 
loop-cluster algorithm, which is very efficient for $B=0$, suffers from 
exponential slowing down for large values of the field. Here we have traded 
this slowing down problem for a severe sign problem by a change of the spin 
quantization axis. Meron-cluster algorithms provide a general strategy to deal 
with sign problems and have previously led to a complete solution of severe 
fermion sign problems \cite{Cha99a,Cha99b,Cha99c,Cox99}. Here we have 
generalized the meron concept to quantum spins in a magnetic field, which again
led to a complete solution of the sign problem for these bosonic systems. Our 
numerical simulations of quantum spin ladders are in agreement with analytic 
predictions and confirm that ladders consisting of an even number of 
transversely coupled spin $1/2$ chains are described by a $(1+1)$-d $O(3)$ 
symmetric quantum field theory.

In conclusion, the meron concept provides us with a powerful algorithmic tool
--- the meron-cluster algorithm --- which can lead to a complete solution of
severe sign problems. In this paper we have demonstrated that this algorithm 
allows us to simulate bosonic quantum spin systems in an arbitrary magnetic 
field. The next challenge is to construct meron-cluster algorithms for systems 
that show high-temperature superconductivity. A first step in this direction 
was taken in \cite{Cox99}. In that paper a Hamiltonian with the symmetries of 
the Hubbard model but with enhanced antiferromagnetic couplings has been
investigated. The enhanced antiferromagnetic couplings allowed us to solve the 
fermion sign problem in that model. It turned out that after doping the model 
undergoes phase separation and does not superconduct. It remains to be seen if 
meron-cluster algorithms can also be constructed for models that show 
high-temperature superconductivity.

\section*{Acknowledgements}

We have benefited from discussions about cluster algorithms with R. Brower, 
H. G. Evertz and M. Troyer. Furthermore, we are grateful to P. Hasenfratz and 
F. Niedermayer for very useful remarks about the Bethe ansatz solution of the 
2-d $O(3)$ model. U.-J. W. also likes to thank the theory group of Erlangen
University where this work was completed for its hospitality and the A. P. 
Sloan foundation for its support.


\begin{thebibliography}{10}

\bibitem{Cha99a}
S. Chandrasekharan and U.-J. Wiese, cond-mat/9902128.

\bibitem{Cha99b}
S. Chandrasekharan, J. Cox, K. Holland and U.-J. Wiese, hep-lat/9906021.

\bibitem{Cha99c}
S. Chandrasekharan, hep-lat/9909007.

\bibitem{Cox99}
J. Cox, C. Gattringer, K. Holland, B. Scarlet and U.-J. Wiese, hep-lat/9909119.

\bibitem{Bie95} 
W. Bietenholz, A. Pochinsky and U.-J. Wiese, Phys. Rev. Lett. 75 (1995) 4524.

\bibitem{Wie92}
U.-J. Wiese and H.-P. Ying, Phys. Lett. A168 (1992) 143.

\bibitem{Eve93}
H. G. Evertz, G. Lana and M. Marcu, Phys. Rev. Lett. 70 (1993) 875.

\bibitem{Wie94}
U.-J. Wiese and H.-P. Ying, Z. Phys. B93 (1994) 147.

\bibitem{Eve97}
H. G. Evertz, The loop algorithm, in Numerical Methods for Lattice Quantum
Many-Body Problems, ed. D. J. Scalapino, Addison-Wesley Longman, Frontiers in
Physics

\bibitem{Bea96}
B. B. Beard and U.-J. Wiese, Phys. Rev. Lett. 77 (1996) 5130.

\bibitem{Pro96}
N. V. Prokof'ev, B. V. Svistunov and I. S. Tupitsyn, 
Pis'ma Zh. Eksp. Teor. Fiz. 64 (1996) 853 [JETP Lett. 64 (1996) 911];
Phys. Lett. A238 (1998) 253; JETP Lett. 87 (1998) 318.

\bibitem{Kas99}
V. A. Kashurnikov, N. V. Prokof'ev, B. V. Svistunov and M. Troyer, Phys. Rev.
B59 (1999) 1162.

\bibitem{San99}
A. W. Sandvik, Phys. Rev. B59 (1999) R14157.

\bibitem{Dag96}
E. Dagotto and T. M. Rice, Science 271 (1996) 618.

\bibitem{Hal83}
F. D. M. Haldane, Phys. Rev. Lett. 50 (1983) 1153.

\bibitem{Khv94}
D. V. Khveshchenko, Phys. Rev. B50 (1994) 380.

\bibitem{Whi94}
S. R. White, R. M. Noack and D. J. Scalapino, Phys. Rev. Lett. 73 (1994) 886.

\bibitem{Sie96}
G. Sierra, J. Phys. A29 (1996) 3299.

\bibitem{Gre96}
M. Greven, R. J. Birgeneau and U.-J. Wiese, Phys. Rev. Lett. 77 (1996) 1865.

\bibitem{Syl97}
O. F. Syljuasen, S. Chakravarty and M. Greven, Phys. Rev. Lett. 78 (1997) 4115.

\bibitem{Cha96}
S. Chakravarty, Phys. Rev. Lett. 77 (1996) 4446.

\bibitem{Bro98}
R. Brower, S. Chandrasekharan and U.-J. Wiese, Physica A261 (1998) 520.

\bibitem{Cha88}
S. Chakravarty, B. I. Halperin and D. R. Nelson, Phys. Rev. Lett. 60 (1988) 
1057; Phys. Rev. B39 (1989) 2344.

\bibitem{Wie85}
P. B. Wiegmann, Phys. Lett. B152 (1985) 209.

\bibitem{Has90}
P. Hasenfratz, M. Maggiore and F. Niedermayer, Phys. Lett. B245 (1990) 522.

\bibitem{Has91}
P. Hasenfratz and F. Niedermayer, Phys. Lett. B268 (1991) 231.

\bibitem{Bea98}
B. B. Beard, R. J. Birgeneau, M. Greven and U.-J. Wiese, Phys. Rev. Lett.
80 (1998) 1742.

\bibitem{Chi97}
R. Chitra and T. Giamarchi, Phys. Rev. B55 (1997) 5816.

\bibitem{Tro98}
M. Troyer and S. Sachdev, Phys. Rev. Lett. 81 (1998) 5418.

\end{thebibliography}
\end{document}